\documentclass{article}
\usepackage{natbib}
\usepackage{graphicx,float,wrapfig}
\usepackage{multirow, threeparttable, color}
\usepackage{hyperref}

\newcommand{\mh}{M_h}
\newcommand{\ms}{M_{\odot}}
\newcommand{\hii}{H II}

\begin{document}
\onecolumn

\title{Simulating the Cosmic Dawn with Enzo} 
\author{Michael L. Norman\footnote{San Diego Supercomputer Center, University of California San Diego, USA} \footnote{Center for Astrophysics and Space Sciences, University of California San Diego, USA } \footnote{Correspondence: mlnorman@ucsd.edu} \and Britton Smith$^*$ \and James Bordner$^*$}

\maketitle

\begin{abstract}
We review two decades of progress using the Enzo hydrodynamic cosmology code to simulate the Cosmic Dawn, a period of roughly 1 billion years beginning with the formation of the first stars in the universe, and ending with cosmic reionization. Using simulations of increasing size and complexity, working up in length and mass scale and to lower redshifts, a connected narrative is built up covering the entire epoch. In the first part of the paper, we draw on results we and our collaborators have achieved using the Enzo cosmological adaptive mesh refinement code. Topics include the formation of Population III stars, the transition to Population II star formation, chemical enrichment, the assembly of the first galaxies, their high redshift galaxy statistics, and their role in reionization. In the second part of the paper we highlight physical difficulties that will require new, more physically complex simulations to address, drawing from a broader literature survey. We discuss the healthy interplay between self-consistent numerical simulations and analytic and semi-analytic approaches. Finally, we discuss technical advances in hardware and software that will enable a new class of more realistic simulations to be carried out on exascale supercomputers in the future.  
\end{abstract}

\section{Introduction}
\label{sec:intro}
The Cosmic Dawn  begins with the formation of the first star in the universe and ends about 1 billion years later when the intergalactic medium has been ionized by UV light from the first galaxies. The later half of Cosmic Dawn has only recently begun to be explored observationally by the Hubble Space Telescope (HST) and other observatories. While much has been learned in recent years \citep{2016ARA&A..54..761S}, many fundamental questions remain unanswered. Our driving question is: how did the universe first light up with stars, and what happened subsequently to make the most distant galaxies seen by the HST? The very earliest phases are beyond the HST's reach, and unfortunately also beyond the reach of the James Webb Space Telescope (JWST), to be launched early in the next decade. Nonetheless a picture has emerged of how structures form and galaxies grow during this tumultuous time through theoretical and computational means. In this article, we present an unbroken narrative connecting the first stars to the first galaxies, ending with cosmic reionization based on our numerical explorations carried out over the past two decades with numerous collaborators. We do this not only to highlight the progress made, but also to highlight the areas of uncertainty that will drive a new generation of simulations using new codes on future high performance computing (HPC) platforms. This is in keeping with the topical theme of this issue,``Imagining the future of astronomy and space science".

There are many excellent reviews of this subject that are not so narrowly focused on numerical simulations, as we are here. For a general theoretical introduction we recommend \cite{BarkanaLoeb2001}. The numerical challenges of simulating the first stars and galaxies, along with a comprehensive survey of results, is given in \cite{Greif2015}. The physics and numerical simulation of reionization is reviewed by \cite{Choudhury2006} and \cite{TracGnedin2011}, respectively. Finally, semi-analytic approaches have been successfully applied to a number of topics discussed below, which we will mention as we go along. 

\subsection{Why simulate the Cosmic Dawn?}
There are two compelling reasons for simulating the Cosmic Dawn. The first is scientific curiosity. It is intrinsically interesting to understand the origins of things, in this case, the first light in the universe. The second is that detailed numerical simulations provide an interpretive basis for understanding what is being observed. This is true of all types of computational astrophysics simulations, but especially true here where the observations are so close to the detection threshold and angular resolution of the instrument. To illustrate this, we show in Fig. 1 a simulation of a dwarf galaxy at $z=15$, its rest frame appearance, and its appearance if observed by the HST and the JWST. We can see that although the simulation provides a lot of detail about the galaxy's morphology and spectrum, this is smoothed over by the instruments. In the lower left corner, show the intrinsic stellar spectrum (thin blue line) and processed galactic spectrum with dust and gas absorption, re-emission, and emission lines (thick green line). The reprocessing of the stellar radiation by the galaxy's interstellar medium is clearly important for a spectroscopic interpretation of observed spectra.

\begin{figure}[h!]
\begin{center}
\includegraphics[width=1.0\columnwidth]{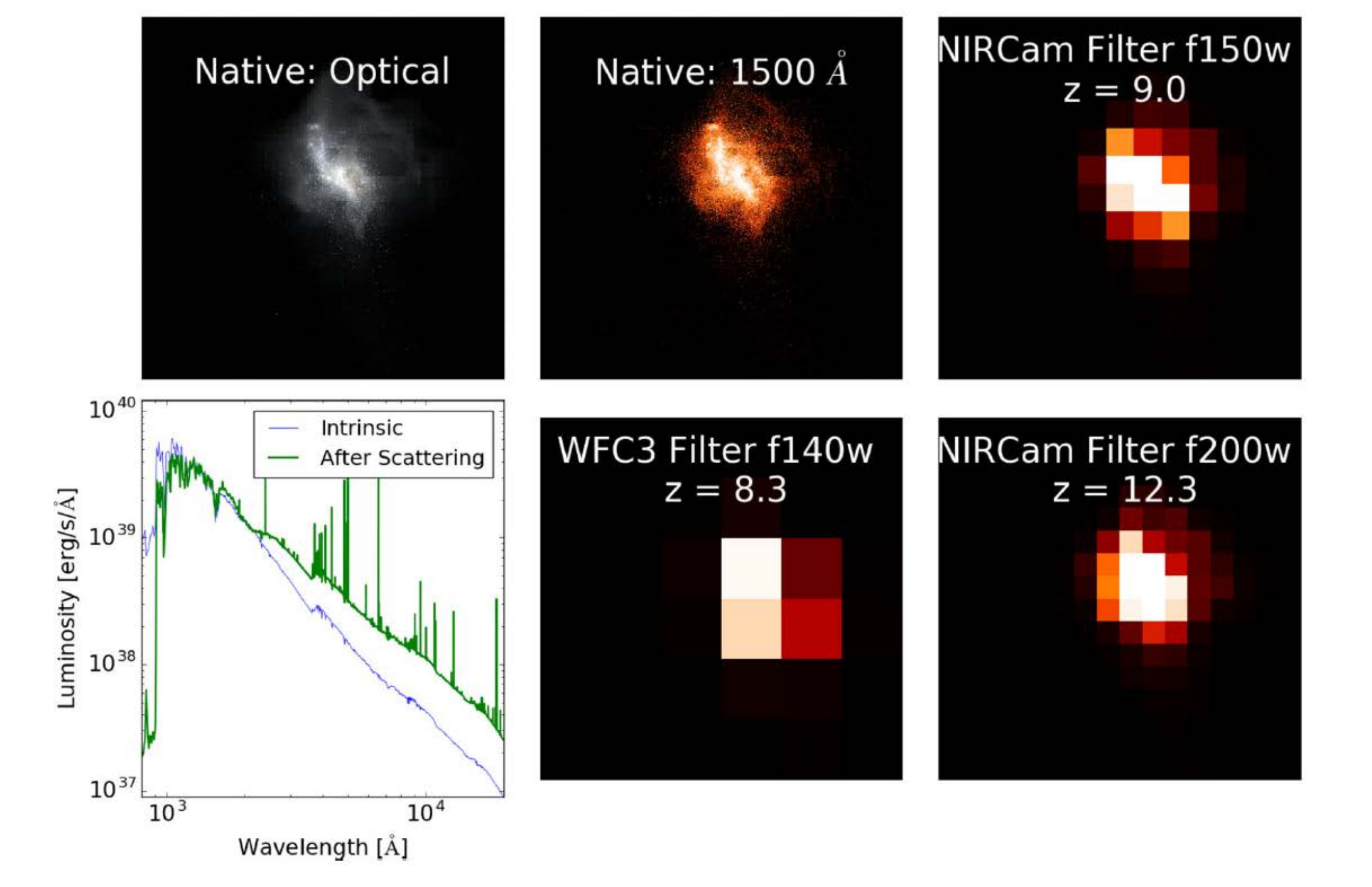} 
\end{center}
\caption{A simulated $z=15$ high redshift protogalaxy and its photometric appearance with Hubble Space Telescope and the James Webb Space Telescope. A partial reproduction of Fig. 12 from \cite{Barrow17a}.}
\label{fig:1}
\end{figure}

\section{Numerical Simulations}
\label{sec:sims}
Most of the simulations discussed in this article were carried out with the Enzo code \citep{Enzo} over the past 18 years. The code, and the supercomputers employed, have both improved enormously over that timeframe. To a real extent, the exponential increase in supercomputer power over the past roughly 2 decades, coupled with Enzo's increasing software capabilities, is what has enabled us to build the narrative. On the hardware side, the empirical ``Moore's law for supercomputers" \citep{Denning2016}, states that peak supercomputer speed roughly doubles every 18 months.  Over 18 years this translates into a speedup of $2^{18/1.5}=4,096$, which is a substantial factor. In fact, over the past half decade, Moore's law for supercomputers has actually sped up due to increased concurrency, making the speedup closer to $10^5$. In astrophysics and cosmology, this increase in capability has
been employed to increase the size and physical realism of the simulations undertaken. 

The Enzo code \footnote{http://enzo-project.org} is a widely used open-source software application that employs adaptive mesh refinement (AMR) to achieve high resolution where needed. Originally developed for numerical cosmology by Greg Bryan from 1994 to 1996 for his PhD thesis at the University of Illinois, Enzo is now a versatile, community-developed application with contributions from numerous authors \citep{Enzo}. Later in this article, we discuss how the open-source software movement was instrumental in the development of Enzo community and code. 

Enzo solves the equations of hydrodynamic cosmology in an expanding universe within a cubic domain of size $L_{box}$. The comoving``boxsize" is chosen to be large enough to encompass the objects of interest, but small enough to be computationally tractable. For simulations of the Cosmic Dawn, $L_{box}$ varies from as small as 1/8 comoving Mpc (cMpc) in the first star simulations to as large as 40 cMpc in the Renaissance Simulations. These are necessarily much smaller than typical boxes used to simulate galaxy large scale structure (100-1000 cMpc) because we are interested in the smallest, earliest objects to form in the universe. 

In its simplest form, Enzo integrates the Euler equations of gas dynamics and the Lagrangian N-body dynamics equations for the dark matter, which are coupled through their mutual self-gravity via the Poisson equation. Enzo's first important innovation was the introduction of structured AMR to numerical cosmology \citep{BryanNorman1997,NB98}(cf. Fig 2), drawing on the the pioneering work of \citet{BergerAMR}. Berger and Colella used structured AMR to improve resolution in problems of shock hydrodynamics. While this is also important in cosmological hydrodynamics, the main purpose for AMR is to maintain adequate numerical resolution in gravitationally collapsing and merging dark matter halos, and later, in gravitationally collapsing protostellar cloud cores \citep{ABN00}. 

The second important innovation, essential for the first star simulations, was the incorporation of a model for primordial gas chemistry including $H_2$, the primary coolant in dense primordial gas. A minimal model was developed by Tom Abel for his Masters thesis for the 9 chemical species $H, H^+, He, He^+, He^{++}, H^-$ $H_2^+, H_2,$ and $e^-$ \citep{Abel97}. A modified backward-Euler numerical method for integrating the time-dependent rate equations for the 9 species was developed by Peter Anninos \citep{Anninos97}. Bryan, Abel, Anninos, and Norman were all together at the National Center for Supercomputing Applications at the University of Illinois in the 1994-1996 timeframe. A synthesis of Bryan's Enzo code, Abel's primordial chemistry model, and Anninos' efficient chemistry solver enabled the first fully cosmological simulations of the formation of the first stars in the universe \citep{ABN00,ABN02}, discussed in more detail below. 

In subsequent years, a variety of additional physics capabilities were incorporated into the Enzo code, driven in large part by the physics of the Cosmic Dawn. Population III (hereafter Pop III) stars are predicted to be massive (Sec. \ref{subsec:pop3}), and as such emit copious amounts of ionizing UV radiation. Therefore, 3D radiative transfer methods were developed and incorporated. The most accurate and flexible method based on adaptive ray tracing was developed by John Wise and Tom Abel \citep{Wise11_Moray}, although an alternative method based on flux-limited diffusion has also proven useful for reionization simulations \citep{Norman15}. Chemical enrichment from Pop III supernovae drives the transition from primordial star formation to Population II star formation. This necessitates the ability for Enzo to track metallicity fields in expanding supernova remnants, and to correctly calculate the radiative cooling due to fine structure lines in a mixture of heavy elements. Since Pop II stars also chemically enrich their environments through winds and supernovae, multiple metallicity fields are needed. John Wise introduced two metallicity fields, one for metals ejected by Pop III supernovae, and one for metals returned to the ISM by Pop II stellar mass loss and supernovae in his ``Birth of a Galaxy" simulations \citep{Wise12_Galaxy}, discussed below. Metal line cooling was introduced into Enzo by Britton Smith, Stein Sigurdsson, and Tom Abel who first explored the transition to metal-enriched star formation with AMR simulations \citep{2009ApJ...691..441S}, discussed in more detail below.

Two other physics ingredients are essential for simulating the Cosmic Dawn: a recipe for star formation and feedback in cases where it is unresolved, as in galaxy formation simulations, and cosmic radiation backgrounds (UV, Xray), that may modify the thermal, chemical, and ionization state of the gas. We defer discussion of these to the relevant sections below. Another piece of physics, magnetic fields, has been implemented in the Enzo code \citep{Collins2010,WangAbel2009}, but so far has not been much explored in the context of the Cosmic Dawn. 

We conclude this section on simulations with a brief discussion about how Enzo is verified and validated. Verification, which asks ``am I solving the equations correctly?", is done at the physics module level through an extensive test suite documented in \cite{Enzo,Wise11_Moray,Norman15}. Many test problems have analytic solutions, which are compared against. Others have reference numerical solutions that are well established. Validation, which asks ``am I solving the correct equations?" is done through comparison with observations as a part of the scientific process, and with the results of other codes through code-comparison campaigns. Enzo has participated in many such campaigns, validating its hydrodynamics \citep{Agertz2007}, magnetohydrodynamics \citep{2011ApJ...737...13K}, cosmological hydrodynamics \citep{SBC99,ON2005}, radiation transport \citep{RT06,Iliev09}, and galaxy formation physics \citep{AGORA}.  

\section{Two Decades of Enzo Results}
\label{sec:results}
\subsection{From primordial gas to Population III stars and remnants}
\label{subsec:pop3}

20 years on from the seminal paper by \citet{Tegmark97}, the story of how primordial gas collects in dark matter minihalos to form the first stars is well known, and the subject of a number of review articles and books \citep{LoebBarkana2001,BrommLarson2004,Glover2005,Norman2008,Greif2015}. As discussed by \citet{Tegmark97}, primordial gas becomes gravitationally bound to dark matter minihalos at high redshifts, and is able to radiatively cool by $H_2$ if there is enough of it. Using an analytical model, they derived a minimum baryonic mass $M_b$ that could cool as a function of redshift. They found that $M_b$ is strongly redshift dependent, dropping from $\sim 10^6 M_{\odot}$ at $z \sim 15$ to $5 \times 10^3 M_{\odot}$ at $z \sim 100$ as molecular cooling becomes effective (multiply by 6 to get halo mass). They estimated ``a fraction $10^{-3}$ of all baryons may have formed luminous objects by $z=30$, which could be sufficient to reheat the universe." 

Tegmark et al.'s results were in good agreement with results from the first fixed grid 3D hydrodynamic cosmological models to incorporate $H_2$ cooling by \citet{AbelEtAl98}. However, neither the analytic model, nor the earliest simulations, could determine what the consequences of that cooling was. That is because star formation occurs on vastly smaller scales than the halo scale, which was only marginally resolved in \citet{AbelEtAl98}. Speculations in the literature concerning what might form ranged from supermassive black holes to clusters of low mass stars. Our current understanding, based on 3D numerical simulations that span a vast range of spatial scales and densities, is that cooling and collapse results in the formation of a single massive star per halo \citep{ABN02,2008Sci...321..669Y}, although massive binary stars may also be produced \citep{2009Sci...325..601T,2010MNRAS.403...45S}.   
In a series of papers beginning in 1998 \citep{ABN98,NB98,NAB99,NAB00,ABN00,BAN01,ABN02}, Tom Abel, Greg Bryan and Mike Norman described numerical AMR simulations of increasingly higher resolution that established the following robust results. 
\begin{itemize}
    \item the analysis of \citet{Tegmark97} is verified numerically, and the minimum mass to cool curve is a good estimate \citep{NAB00};
    \item $H_2$ cooling produces a gravitationally unstable core of mass $\sim 1000 M_{\odot}$, density of $\sim 10^4$ cm$^{-3}$ and temperature $\sim 200K$; 
    \item once the molecular hydrogen fraction in the core exceeds $f_{H_2} \sim 5 \times 10^{-4}$, it collapses rapidly in an inside-out fashion;
    \item collapse produces a fully molecular hydrogen core of less than a solar mass that accretes at a rate of $\sim 10^{-3} M_{\odot}$/yr; 
    \item the core does not fragment due to chemo-thermal instability to the highest densities achieved $n \sim 10^{12}$ cm$^{-3}$.
\end{itemize}

These findings were independently verified by \cite{Bromm02_P3} and following work using smoothed particle hydrodynamics (SPH) simulations. The implication of the last two findings is that, absent fragmentation or radiation feedback shutting off accretion \citep{McKeeTan08}, the final star will be massive, as it would accrete $\sim 100 M_{\odot}$ in a Helmholtz time of $10^5$ yr \citep{ABN02}. The first simulation to form a hydrostatic protostar starting from cosmological initial conditions was the SPH simulation of \cite{2008Sci...321..669Y}. They found a single protostellar seed of 1\% the Sun's mass is formed in the center of the parent minihalo, and speculated that it would grow to become a massive star. Other authors \citep{2011Sci...331.1040C,2012MNRAS.424..399G} have simulated a short way into the disk accretion phase of the protostar, and found that the disk fragments into numerous smaller accreting bodies. We return to the issue of fragmentation in a later section when we discuss future simulations. 

The final masses of primordial stars is fundamental to how the Cosmic Dawn develops next. If the stars are massive, as simulations indicate, they will be very luminous in the UV, live only a few million years, chemically enrich the universe through their supernova explosions, and leave behind stellar remnants (neutron stars and black holes) \citep{2002A&A...382...28S,2002ApJ...567..532H}. This is the conventional wisdom, and the scenario most explored by us and others. Others have argued that the universe may contain low-mass primordial stars arising from primordial protostellar disk fragmentation \citep{2010MNRAS.403...45S,2011ApJ...737...75G}. We know that at lease {\em some} primordial stars must have been massive enough to create the first heavy elements of the periodic table, as we have no other formation mechanism available. In the narrative that follows, we assume the conventional wisdom that primordial stars are {\em primarily} massive, but do not require that they are {\em only} massive. 

A key complication that arises if primordial stars are primarily massive, is that they create a {\em negative feedback effect} on their subsequent formation as a result of the $H_2$ dissociating radiation they emit in the Lyman-Werner band $10.2-13.6$ eV. An extensive early population of massive primordial stars builds up a Lyman-Werner background (LWB) radiation field, that can suppress or limit subsequent primordial star formation by photodissociating the $H_2$ molecule whose presence is required to form them in the first place. Using Enzo, \cite{Machacek01} found that the minimum halo mass capable of cooling by $H_2$ increases with the mean intensity of the LWB, and above a certain intensity, halos as massive has $\mh=10^7 M_{\odot}$ are unable to cool.  The first numerical+semianalytic study to attempt to self-consistently model negative feedback in a cosmological setting was by \cite{yoshida03}. They found that the global Pop III star formation rate becomes self-regulating at $z\approx 30$ as minihalos below an increasing minimum mass to cool become suppressed. However, \cite{OShea08} showed, using high resolution Enzo AMR simulations, that even for very high LWB intensities, $H_2$ formation and cooling is not suppressed, but merely delayed until halo virial temperatures increase enough so that $H_2$ formation exceeds destruction. This occurs in halos of mass exceeding a few $\times 10^7 M_{\odot}$. 

\begin{figure}[h!]
\begin{center}
\includegraphics[width=0.8\columnwidth]{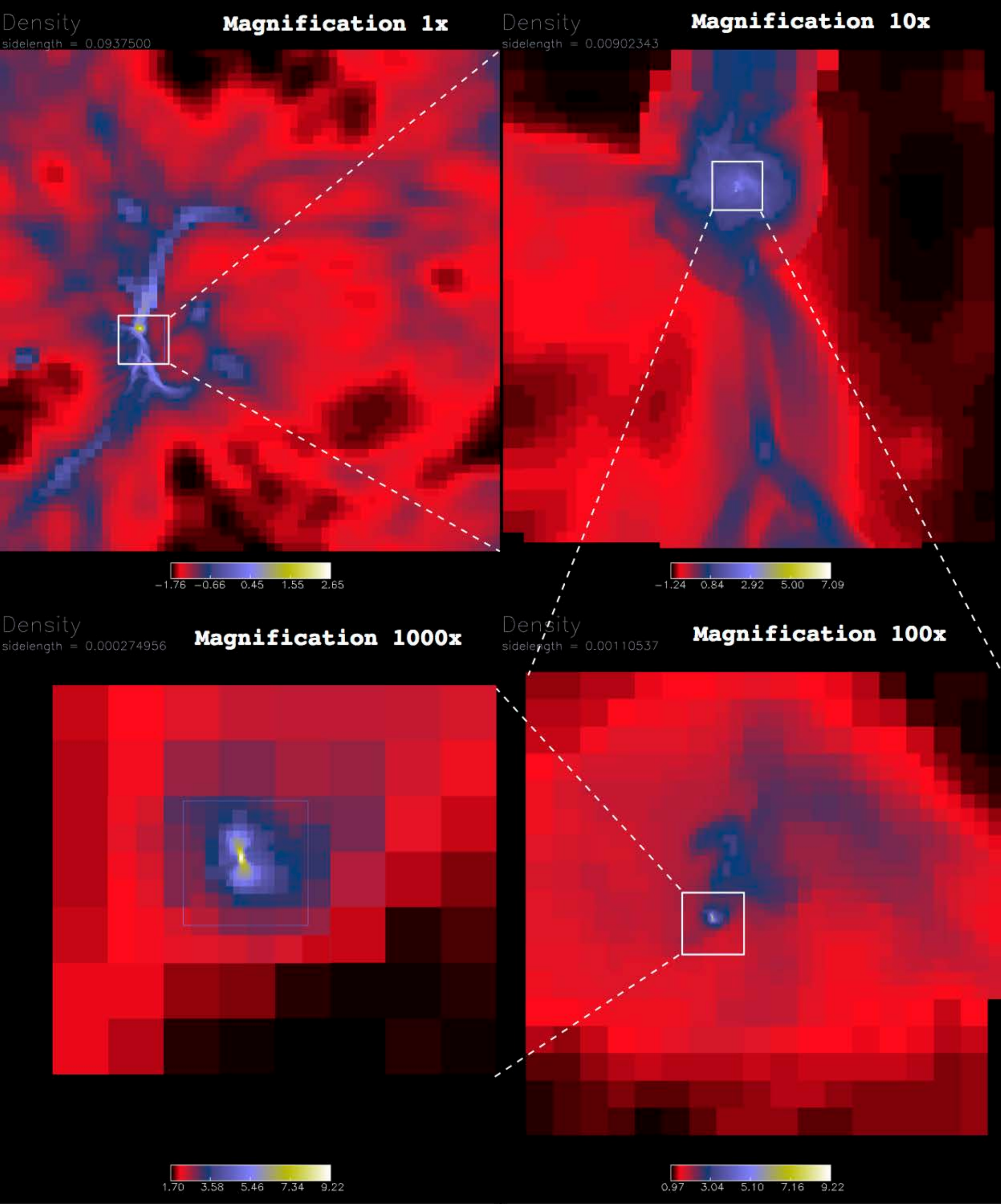}
\end{center}
\caption{An early zoom-in image from a cosmological AMR simulation of primordial star formation. Reproduced by permission of Figure 1 from \cite{NB98}.}
\label{fig:2}
\end{figure}

\subsection{Pop II star formation in the ashes of Pop III supernovae}
\label{subsec:pop2}
The conclusion that Pop III stars were massive, based on the simulations discussed above, stands in contrast to what we know about metal-enriched stars in our Galaxy. For any observed group of stars, the distribution of stellar masses (the number of stars as a function of mass) follows a power-law with a slope of roughly -2.35, a quality first reported more than 60 years ago by \citet{1955ApJ...121..161S}. Below about a solar mass, the distribution function turns over and declines \citep{Kroupa1993, Chabrier03}. In other words, there are far more low mass stars than high mass.  This appears to hold almost universally over a broad range of physical conditions \citep{Kroupa2012}.  This is itself very different from the case of Pop III, where even small changes to the properties of host halos result in stellar masses that can vary by up to two orders of magnitude \citep{2014ApJ...781...60H}.  How and when did this transition in star formation take place?

\begin{figure}[h!]
\begin{center}
\includegraphics[width=10cm]{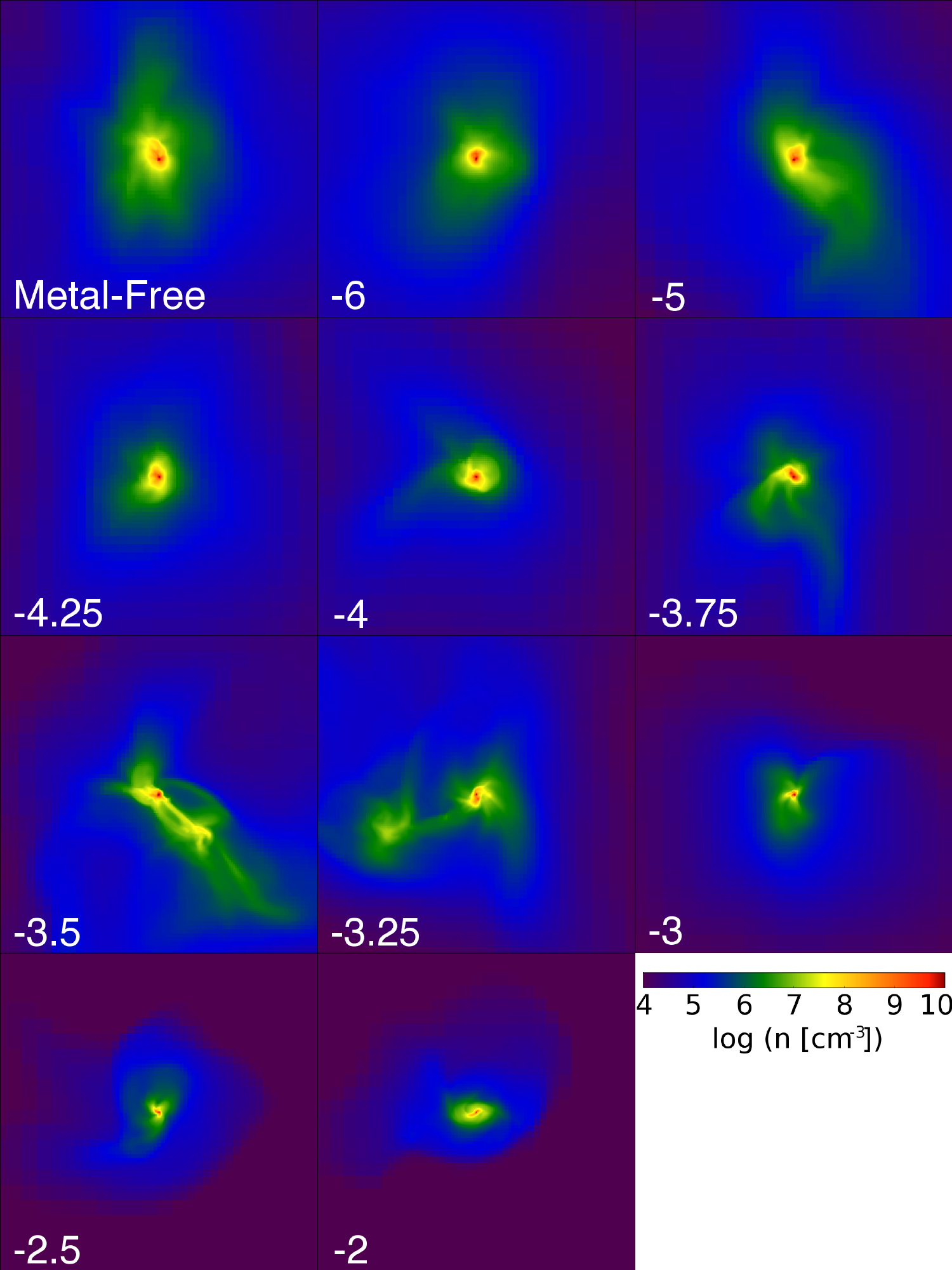}
\end{center}
\caption{Density projections of the central 0.5 pc in collapsing gas clouds at different metallicities from simulations using Enzo.  At metallicities above 10$^{-3.5} Z_{\odot}$, the cloud is able to fragment due to enhanced cooling from metals.  Below 10$^{-3.5} Z_{\odot}$, the gas reaches the CMB temperature floor at an early stage and becomes stabilized to fragmentation.  Reproduced by permission of the American Astronomical Society, Figure 1 from \cite{2009ApJ...691..441S}.}
\label{fig:Smith08}
\end{figure}

In their supernova explosions, Pop III stars introduced the first heavy elements (metals) to the universe.  The addition of metals to a gas adds to the number of atomic and molecular transitions, increasing its ability to cool.  Recalling that, to first order, Pop III stars are massive because metal-free gas clouds cease to cool (and, thus, fragment) as they collapse, one can then calculate the additional metals required for cooling and fragmentation to continue.  From this exercise, two "critical metallicities" can be identified, at $\sim 10^{-3.5} Z_{\odot}$ due to fine-structure lines from C and O \citep{2003Natur.425..812B} and at $\sim 10^{-5.5} Z_{\odot}$ due to thermal emission from dust grains \citep{2000ApJ...534..809O}.  Numerous studies using three-dimensional simulations of homogeneously enriched gas clouds have shown that fragmentation does occur at these critical metallicities \citep{bromm01, 2007ApJ...661L...5S, clark08, 2009ApJ...691..441S, 2011ApJ...729L...3D}.  For example, Figure \ref{fig:Smith08} shows density projections from Enzo simulations of gas clouds at different metallicities from \citet{2009ApJ...691..441S}.  These simulations did not include dust, but the change in morphology at $Z = 10^{-3.5} Z_{\odot}$ is evident.  Additionally, \citet{2009ApJ...691..441S} found that fragmentation was again reduced at even higher metallicities ($Z > 10^{-3} Z_{\odot}$) as the gas was stabilized by reaching the temperature floor of the high-redshift CMB (almost 60 K at $z \sim 20$).  However, other simulations claimed that the ability of gas to fragment was extremely dependent on the initial setup of the simulation \citep{2009ApJ...696.1065J, 2009ApJ...694.1161J, 2014ApJ...783...75M}.  For example, simulations without rotation or turbulence in the cloud show no fragmentation at any metallicity.  The only way to resolve this is to understand the true initial conditions of the first low mass stars.

Addressing this issue is an extreme technical challenge.  In place of simulations of gas at constant metallicities, we must now consider metals ejected by Pop III supernovae and their subsequent mixing into the surrounding medium.  However, even before the explosion, the intense radiation emitted during the main-sequence lifetime of a Pop III star will remove most of the gas from its host halo \citep{whalen04, Kitayama05, whalen08}, in essence plowing the road before the supernova blast-wave sweeps through.  This makes radiative transfer crucial to this effort.  Additionally, a number of criteria conspire to place demands of extraordinary mass and spatial resolution on this type of simulation.  The dark matter mass resolution must be high enough to resolve the smallest Pop III-forming minihalos, but the volume must also be large enough to contain a number of these halos.  As well, extreme adaptive mesh-refinement is necessary in order to follow the collapse of metal-enriched gas to high densities.

\citet{2015MNRAS.452.2822S} presented \textit{PopIIPrime}, the first simulation to begin from cosmological initial conditions, follow the creation and dispersal of metals by Pop III supernova, and end with the formation of metal-enriched stars.  Run on the Blue Waters sustained petascale HPC system at the National Center for Supercomputing Applications, University of Illinois Urbana-Champaign, \textit{PopIIPrime} combined Pop III star formation, radiation transport, and sophisticated chemistry ($H_{2}$ chemistry with metal cooling and dust-grain effects) with a dark matter mass resolution of 1 $M_{\odot}$ and a maximum AMR spatial resolution of nearly 1 AU.  At its onset, the outcome of this enormous effort was entirely unknown.  Would metal-enriched star formation have to wait for the heavy elements to fall back into their halos of origin?  With minihalos, such a process can take up 100-200 Myr or more \citep{Greif10, Jeon14}.  Would metals have to be pulled in during the hierarchical assembly of larger halos \citep{Greif07, Wise12_Galaxy}?

\begin{figure}[h!]
\begin{center}
\includegraphics[width=1.0\columnwidth]{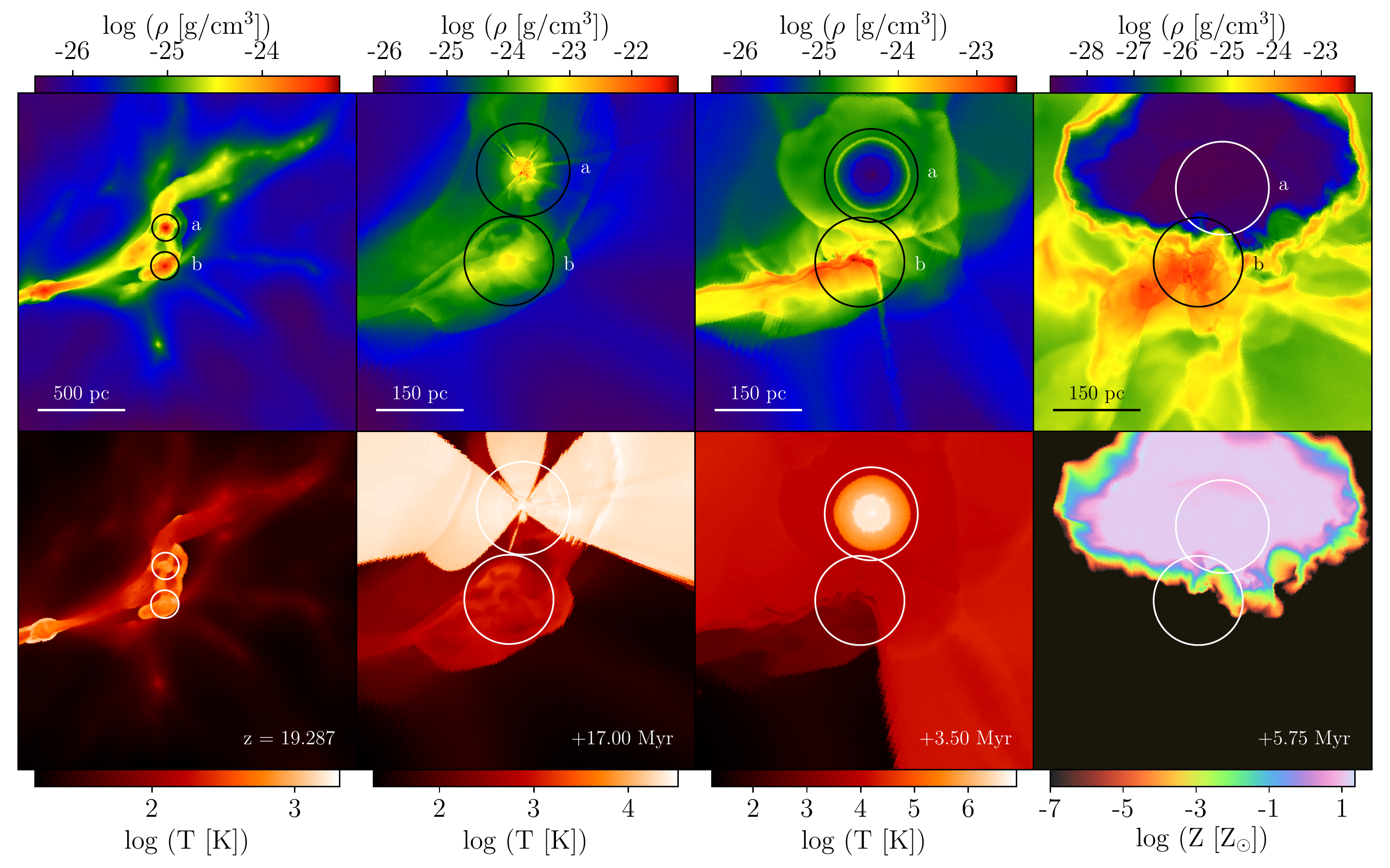}
\end{center}
\caption{The external enrichment of a starless minihalo (halo b) by a Pop III supernova from a nearby minihalo (halo a). The top row shows slices of the gas density and the bottom shows gas temperature and metallicity in the right-most panel. Circles denote the virial radii of the two halos. Column 1: the two halos are shown at z$\sim19$, shortly before the formation of a Pop III star in halo a. Column 2: the star forms in halo a and ionizing radiation immediately escapes. Column 3: the star explodes in a core-collapse supernova. Radiation from the star has compressed the gas in halo b and its encompassing filament. Column 4: the metal-rich blast-wave impacts halo b. Turbulent motion will enable the metals to mix into the densest gas within halo b. A reproduction of Figure 1 from \cite{2015MNRAS.452.2822S}.}
\label{fig:Smith15enrich}
\end{figure}

\begin{figure}[h!]
\begin{center}
\includegraphics[width=10cm]{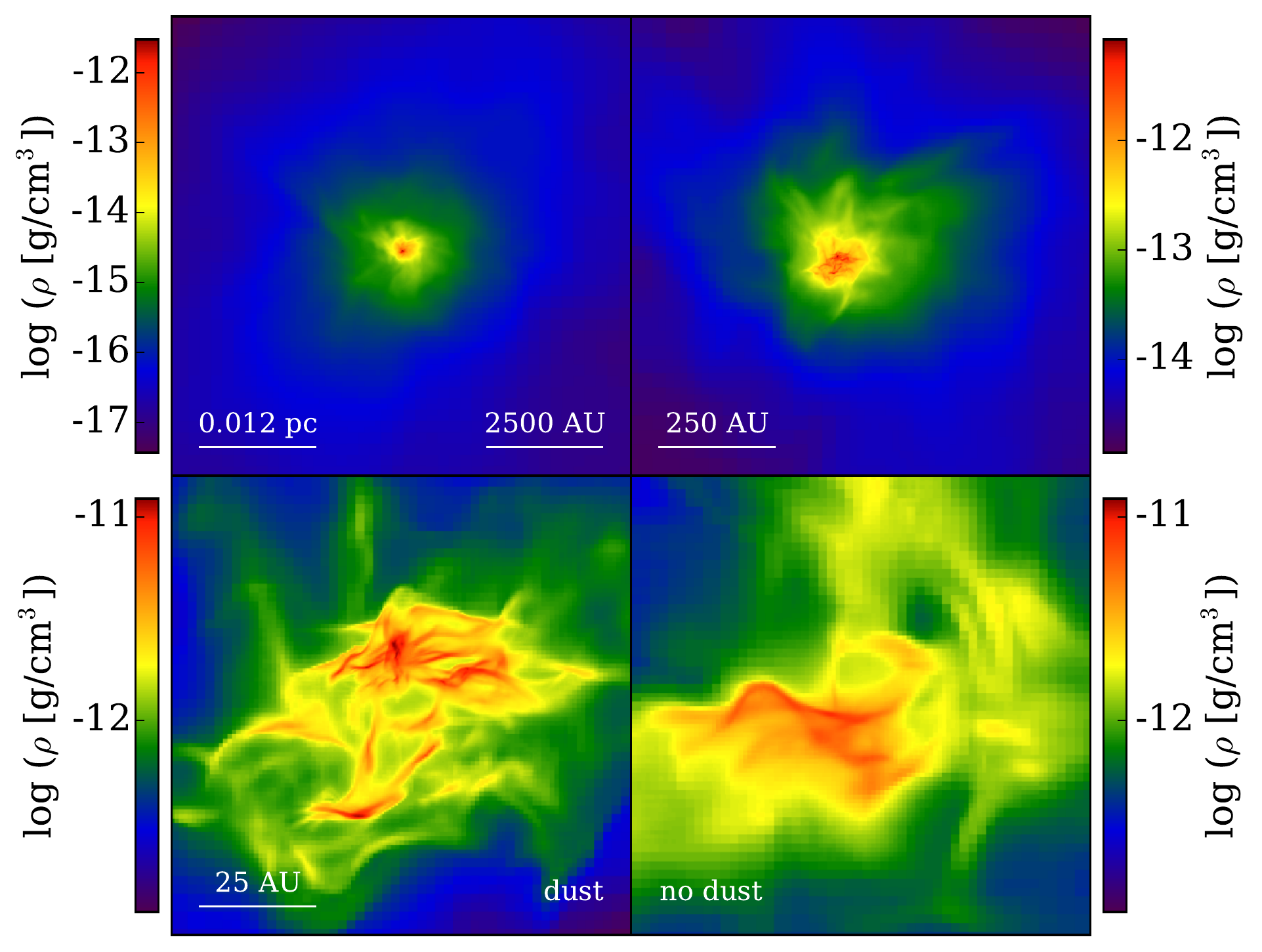}
\end{center}
\caption{The collapse of metal-enriched gas from scales of 2000 AU down to 100 AU, where dust cooling cause the cloud to fragment.  The bottom-right panel shows the result of the same collapse without the effects of dust.  A reproduction of Figure 6 from \cite{2015MNRAS.452.2822S}.}
\label{fig:Smith15collapse}
\end{figure}

Somewhat surprisingly, metal-enriched star formation was found to occur only 25 Myr after the second Pop III supernova, at a redshift of roughly 16.5.  Figure \ref{fig:Smith15enrich} illustrates the unexpected sequence of events.  At $z \sim 19$, a Pop III star forms in the minihalo denoted by circle "a" in the top, left panel of Figure \ref{fig:Smith15enrich}.  Circle "b" denotes a halo of nearly equal mass approximately 200 pc away.  Columns 2 and 3 show the radiative feedback and initial explosion.  The neighboring halo has been heavily disrupted by radiation from the star.  Within 6 Myr of the initial explosion, the blast wave impacts the neighboring halo.  The turbulent motions produced from a combination of the halo's own relaxation and the collision allow metals to mix into the densest gas in the halo's core.  As the density continues to rise, the collapse time scale becomes shorter than the turbulent mixing time scale, and a runaway collapse begins within a pocket of gas enriched to a metallicity of about $2\times10^{-5} Z_{\odot}$.  The collapse of this gas is shown in Figure \ref{fig:Smith15collapse}.  Because the low metallicity, the initial phases of collapse show no fragmentation.  Eventually, at densities of roughly $10^{12} cm^{-3}$, dust cooling triggers vigorous fragmentation on scales of $\sim100$ AU, as shown in the bottom-left panel.  To confirm that dust is the instigator of this, the collapse portion of the simulation is run again without the effects of dust, with the result shown in the bottom-left panel.  This result showed the validity of a so-called "external enrichment mechanism" as a likely scenario for the formation of the earliest stars resembling those that are observed today.  It also highlighted the need to continue to drive simulations to higher and higher resolution and sophistication.

\subsection{Birth of a Galaxy}
We learned in Sec. \ref{subsec:pop3} how the universe converts primordial gas into a primordial star in the cores of dark matter minihalos. And we have seen in Sec. \ref{subsec:pop2}  how the heavy elements in the supernova ejecta of a primordial star can pollute neighboring minihalos, resulting in the formation of a small cluster of metal-enriched stars in some of them. Now we turn to the formation of the first objects in the universe that can rightfully be called galaxies. These are high redshift, dwarf galaxies like the one shown in Fig. \ref{fig:1}, composed of dark matter, metal-enriched gas, and stars. Ideally, we would like a numerical simulation that starts with linear matter fluctuations, self-consistently computes the formation of Pop III stars in their minihalos, as well as their feedback effects, computes the transition to Pop II star formation in chemically enriched gas, and produces one or more protogalaxies automatically. In the {\em Birth of a Galaxy} paper series (hereafter BOG), John Wise and his collaborators used an enhanced version of Enzo to accomplish that. Here we summarize some of their key findings. 

\begin{figure}[h!]
\begin{center}
\includegraphics[width=1.0\columnwidth]{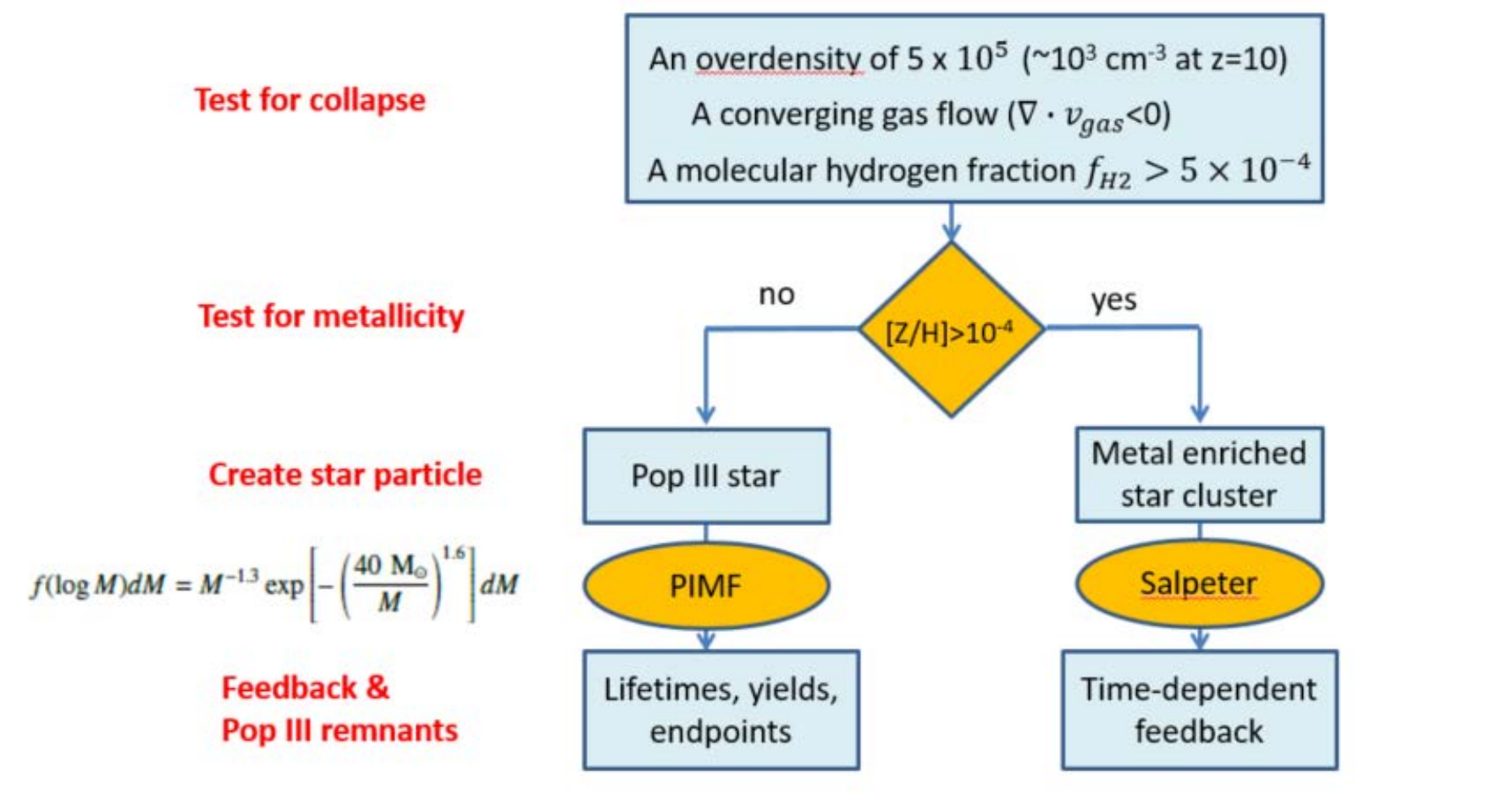}
\end{center}
\caption{Recipe for making a numerical star or star cluster used in Enzo for Cosmic Dawn simulations. Based on description in \cite{Wise12_Galaxy}.}
\label{fig:3}
\end{figure}

Before we do, we must discuss a technical issue. A fundamental requirement of such simulations is a subgrid model for forming stars in regions of cool dense gas that are conducive to star formation, since we lack the resolution to form stars directly. Such models can be viewed as recipes that build in our knowledge about how star formation and stellar evolution works from both observations and theoretical models. \cite{Wise12_Galaxy} introduced the detailed model shown in Fig. \ref{fig:3}. It handles both Pop III and metal-enriched star formation and feedback, extending an earlier model which only handled Pop III star formation \citep{Wise08_Reion}. It consists of a local test for collapse, building in criteria determined from Pop III star formation simulations discussed above, and two conditional branches depending on the value of the local gas metallicity. If $Z/Z_{\odot}<10^{-4}$, a star particle is created which represents an individual Pop III star with a mass drawn from a parameterized primordial initial mass function (PIMF). If $Z/Z_{\odot} > 10^{-4}$, a star particle is created which represents a metal-enriched star cluster of mass $10^3 M_{\odot}$, with an assumed Salpeter IMF. Stellar lifetimes, and their radiative, chemical, and kinetic feedbacks are all determined from the best available stellar models. Depending on the mass assigned to a Pop III star from the PIMF, its endpoint may be either a Type II supernova, a pair instability supernova (PISN), or a prompt black hole without an explosion according to the models of \citep{2002ApJ...567..532H}. We mention this level of detail because it creates variety in the formation history of the first galaxies.
\bigskip

The focus of the first {\em Birth of a Galaxy} paper \citep{Wise12_Galaxy} is the evolution of the stellar populations in a small sample of high redshift dwarf galaxies $10^7 < \mh/M_{\odot} < 10^9$ taking into account chemical enrichment from Pop III supernovae. Achieving a maximum spatial resolution of 1 comoving parsec within a 1 cMpc box, individual molecular clouds were well resolved, and the history of star formation and chemical evolution could be simulated in detail. The entire simulation contained 38 galaxies with 3640 Pop II stellar clusters and captured the formation of 333 Pop III stars. The supernova blasts from Pop III stars first sweep baryons out of their parent minihalos. However the gas fraction recovers from 5\% to nearly the cosmic fraction in halos above $10^7 M_{\odot}$. The simulation showed that a single pair instability supernova is sufficient to enrich the host halo to a metallicity floor of $10^{-3} Z_{\odot}$, triggering a transition to Population II star formation. Despite the small sample of galaxies, their individual star formation and chemical histories showed considerable variation. They found that stellar metallicities do not necessarily trace stellar ages, as mergers of halos with established stellar populations can create superpositions in the age-metallicity stellar tracks. 

The second {\em Birth of a Galaxy} paper \citep{Wise12_RP} examined the role of radiation pressure on star formation in high-redshift dwarf galaxies, and found that it is significant. It was found that momentum input on neutral gas from ionizing radiation from young, massive stars raises the turbulent velocities in star forming clouds compared to a simulation that omits radiation pressure. The heightened level of turbulence is found to significantly reduce the star formation rate and lower the mean stellar metallicity, consistent with observed dwarfs in the Local Group. Radiation pressure is also found to help drive dense gas away from star forming regions, so that supernovae occur in a warmer and more diffuse medium. This launches metal-rich outflows and avoids the ``overcooling problem" seen in some simulations. 

\begin{figure}[h!]
\begin{center}
\includegraphics[width=1.0\columnwidth]{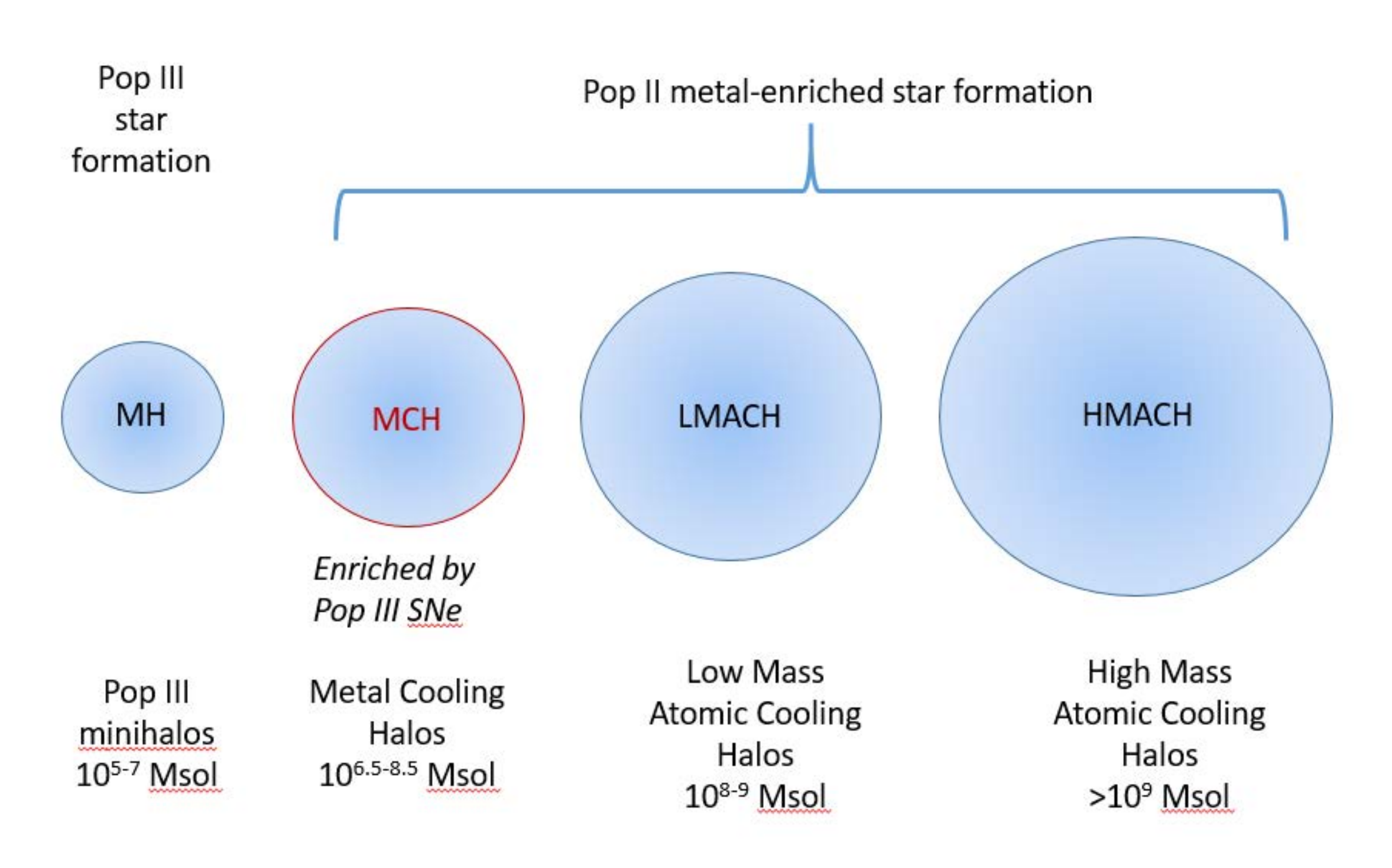}
\end{center}
\caption{Four types of star-forming halos of importance during the Cosmic Dawn. The MHs, LMACHs, and HMACHs have been well-discussed in the literature. The MCHs, discovered in the {\em Birth of a Galaxy} simulations, represent a new, potentially important class.}
\label{fig:4}
\end{figure}

The third {\em Birth of a Galaxy} paper \citep{Wise14} examined the contribution of the smallest galaxies to the process of cosmic reionization. The conventional wisdom is that galaxies in dark matter halos below $\sim 10^8 M_{\odot}$ contribute negligibly to cosmic reionization because they are unable to cool by atomic hydrogen lines and consequently they form stars very inefficiently. Our simulations revealed a new class of low mass halo, dubbed metal cooling halos (MCHs), with masses $10^{6.5} < \mh/M_{\odot} < 10^{8.5}$, that have been chemically enriched by earlier Pop III supernovae (Fig. \ref{fig:4}). Although the MCHs form stars inefficiently, they are numerous. Moreover, we find that a high fraction of the ionizing radiation they produce escapes the host halo, decreasing from 50 to 5 percent in the mass range $10^7 < \mh/M_{\odot} < 10^{8.5}$ . The combination of high space density and high ionizing escape fraction means that the MCHs contribute significantly to reionization, especially in the early stages of reionization. This possibility is further examined below.

\subsection{The high redshift galaxy population: the {\em Renaissance Simulations}}
\label{subsec:rs}
So far, we have discussed phases of the Cosmic Dawn that are beyond the reach of our most powerful space telescopes. However, as the HST has probed ever-further into the past, and as numerical simulations deployed on ever more powerful supercomputers push to lower redshifts, they meet in the middle. Where they meet is the population of high redshift galaxies observed by the HST, at redshifts of $z=6-10$. \cite{2016ARA&A..54..761S} provides an excellent summary of the observations to date. The James Webb Space Telescope is poised to revolutionize this area in the coming years. 

\begin{figure}[h!]
\begin{center}
\includegraphics[width=10cm]{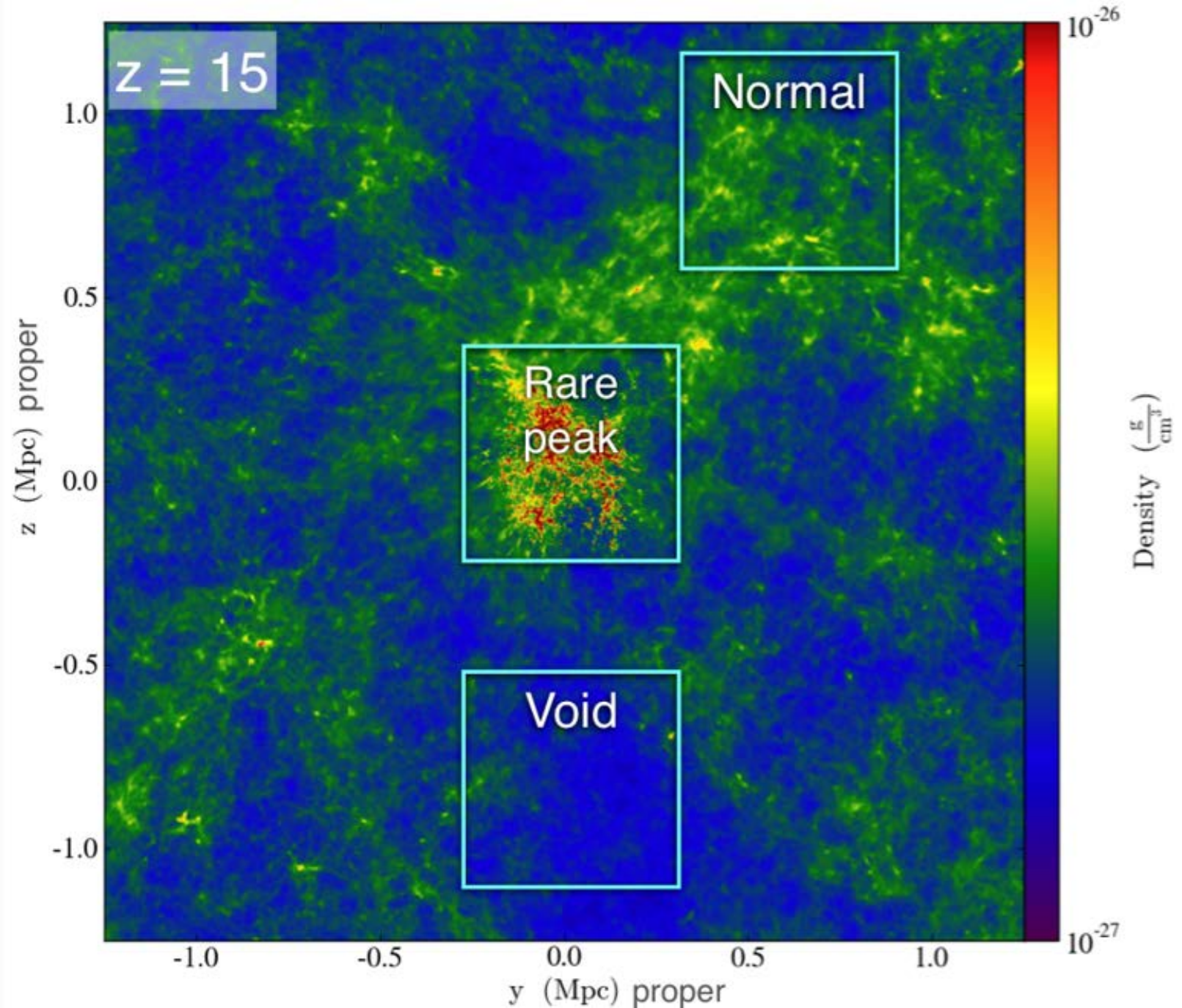}
\end{center}
\caption{The {\em Renaissance Simulations} calculate star and galaxy formation in three regions with AMR and the full physics suite of the {\em Birth of a Galaxy} simulations: a high density region ``Rare peak", a region of average density ``Normal", and a low density region ``Void". The box size is 40 cMpc on a side, which at $z=15$ is a few proper Mpc on a side.}
\label{fig:5}
\end{figure}

The {\em Birth of a Galaxy} simulations lack the statistics to adequately predict the properties of the observed high redshift galaxy population due to the small volume simulated (1 cMpc)$^3$. As emphasized by \citet{BarkanaLoeb2004}, such small volumes do not adequately sample the variance in the dark matter density field, resulting in an understimation of the formation redshifts of Pop III stars and galaxies in rare peaks and their clustering space. All the simulations discussed above have been carried out in very small volumes, and therefore suffer from these defects. For this reason, we have not been very precise about {\em when} certain phases of the Cosmic Dawn begin and end. To rectify these defects, in 2013 we began a new series of simulations called the {\em Renaissance Simulations} which are similar to the {\em Birth of a Galaxy} simulations but in a much larger volume--(40 cMpc)$^3$. Rather than attempt to perform high resolution AMR throughout the entire volume, which would have been computationally prohibitive, we selected three regions of varing mean density for selective refinement (cf. Fig. \ref{fig:5}). The total volume of the 3 refinement regions is about 576 cMpc$^3$--576 times the volume of the BOG simulations. The ``Rare peak", ``Normal" and ``Void" simulations were run to final redshifts of 15, 12.5, and 7.6, respectively. At their final redshifts, the three simulations produced a sample of nearly $10,000$ high redshift galaxies in the mass range $10^7 < \mh/M_{\odot} < 10^8$, nearly 500 galaxies in the mass range $10^8 < \mh/M_{\odot} < 10^9$, and 12 galaxies with $\mh/M_{\odot} > 10^9$. With this sample, we can begin to characterize the statistical properties of high redshift galaxies, and discuss their formation redshifts.  

The {\em Renaissance Simulations} represent the most comprehensive attempt yet to numerically simulate the detailed transition from a starless universe to one filled with galaxies. We have published 10 papers analyzing different aspects of the Renaissance Simulations \citep{Xu13,Xu14,Chen14,Ahn15,OShea15,Xu16a,Xu16b,Xu16c,Barrow17a,Barrow17b}. Here we highlight just a few of the more interesting findings that start to fill in the Cosmic Dawn narrative. 

\medskip
\noindent
\underline{Denser regions get a head start}

\begin{figure}[h!]
\begin{center}
\includegraphics[width=9cm]{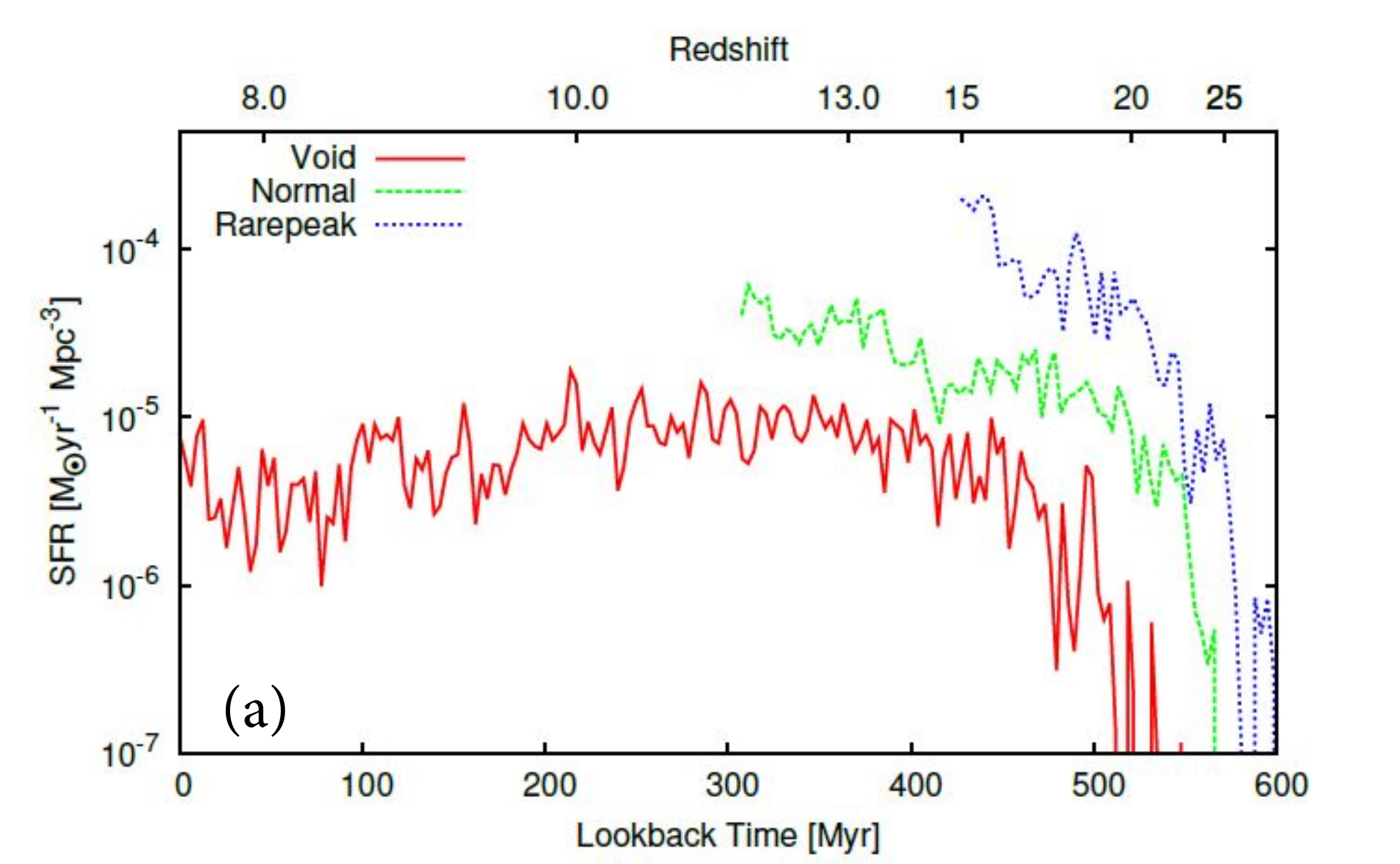}
\includegraphics[width=9cm]{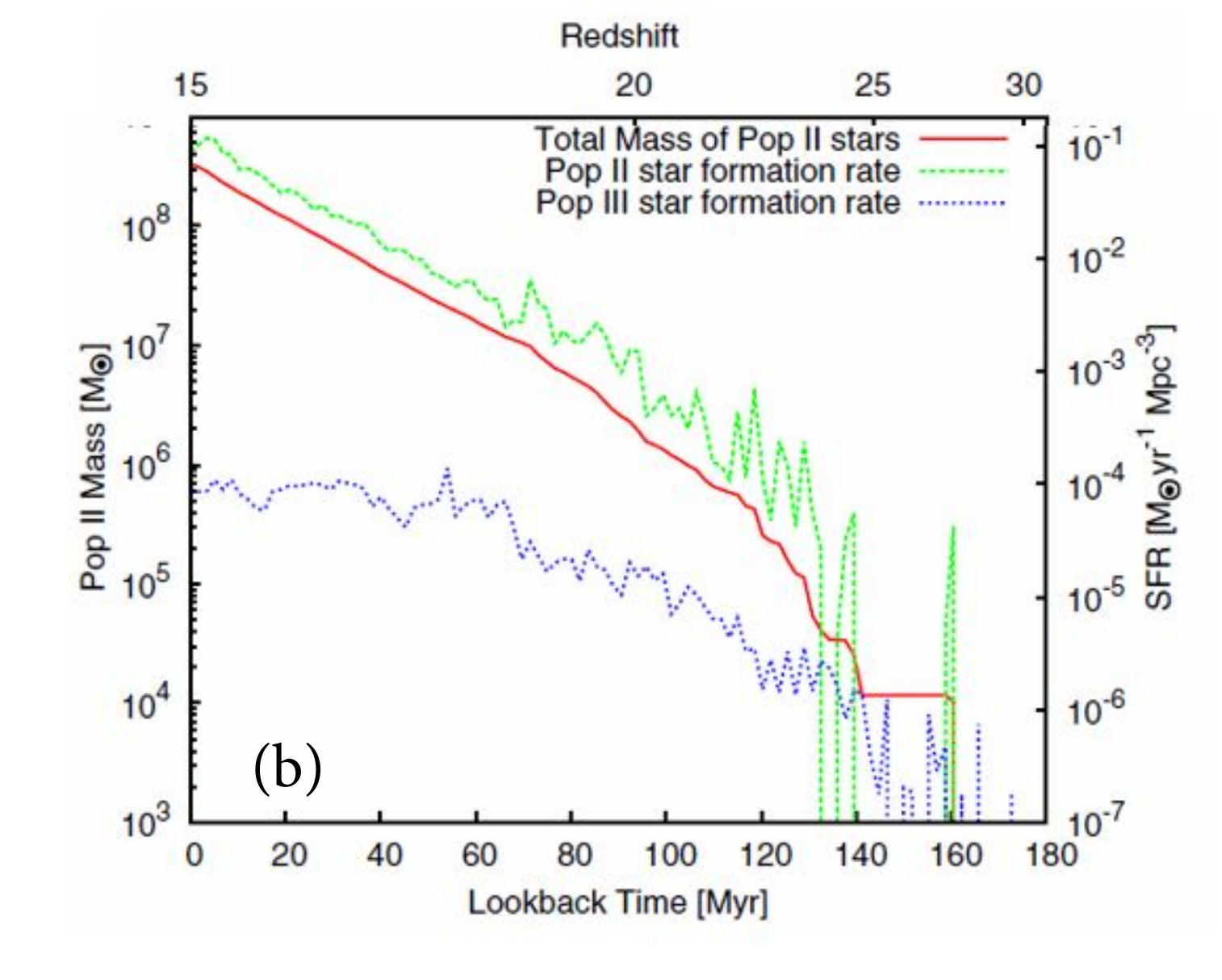}
\end{center}
\caption{Star formation histories in the {\em Renaissance Simulations}. Lookback time is respect to the end of the simulation. Evolutionary time increases from right to left. (a): The volume-averaged Population III star formation rate density in Rarepeak, Normal, and Void regions. Due to the presence of a strong Lyman-Werner background, the onset of Pop III star formation is delayed, but begins earlier in denser regions. The curves end when the simulations were terminated. Reproduced by permission of the American Astronomical Society, Figure 1 from \cite{Xu16c}. (b): Pop III and Pop II star formation rates in the Rarepeak simulation. Pop II quickly overtakes Pop III. Reproduced by permission of the American Astronomical Society, Figure 2 from  \cite{Xu13}. }
\label{fig:6}
\end{figure}

While this is well known from pure dark matter simulations and theory \citep{BarkanaLoeb2004}, Fig. \ref{fig:6}a shows this graphically. Plotted is the volume averaged Pop III star formation rate in the 3 regions. Primordial star formation starts earlier in the denser {\em Rare peak} simulation than in a region of average density, which in turn starts earlier that in a region of low density. The rates grow rapidly and then level out to rates that order themselves in the same way. 

\medskip
\noindent
\underline{The transition from Pop III to Pop II star formation is quick} 

Chemical enrichment from Pop III supernovae triggers the transition to Pop II star formation. As found in the BOG simulations, a single pair instability supernova is sufficient to raise a halo's metallicity above the critical threshold. In each of the three regions, the volume-averaged Pop II star formation rate begins to exceed the Pop III star formation rate within 20-30 Myr after the first Pop III star forms. Thereafter it grows quickly, and exceeds the Pop III star formation rate density by orders of magnitude (Fig. \ref{fig:6}b).  

\medskip
\noindent
\underline{Pop III star formation is not entirely extinguished}

Interestingly, Pop III stars continue to form at low rates in each of the simulations as far as we can advance them. Our longest running simulation, the Void simulation, produces Pop III stars as late as $z=7.6$ \citep{Xu16a}. This is possible because chemical enrichment is local and slow, leaving more than 80\% of the volume pristine by the end of the simulations.  

\medskip
\noindent
\underline{Pop III galaxies should exist}

A total of 14 Pop III galaxies are found in the Void simulation at $z=7.6$ \citep{Xu16a}. Some are found to have multiple active Pop III stars within them. This is a consequence of the fact that Pop III star formation is suppressed by the strong Lyman-Werner background until the halo becomes massive enough. Once the halo masses exceed about $10^7 M_{\odot}$, as much as $10^3 M_{\odot}$ pristine gas is able to cool and form multiple Pop III stars according to our recipe (Fig. \ref{fig:3}). The observational signatures of such galaxies has been analyzed by \cite{Barrow17b}.

\medskip
\noindent
\underline{The Pop II galaxy luminosity function flattens at the faint end}

HST observations of high redshift galaxies probe the brightest end of the luminosity function, whereas our simulations predict the faint end. Where they overlap, the agreement is good. However, at fainter magnitudes the ultraviolet luminosity function (ULF) flattens out and even drops, as it samples smaller halos incapable of forming stars efficiently and continuously \citep{OShea15}.

\begin{figure}[h!]
\begin{center}
\includegraphics[width=0.8\columnwidth]{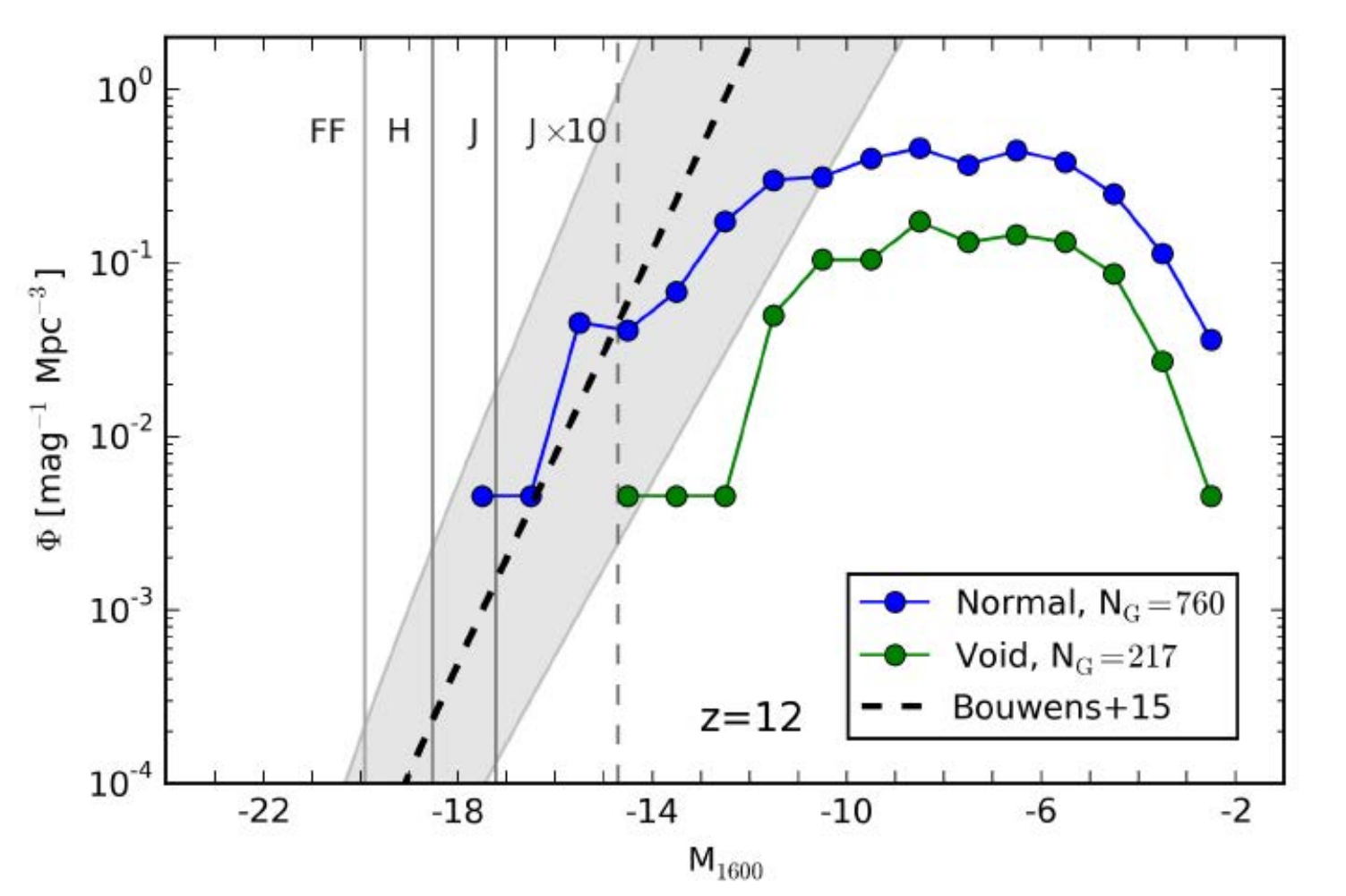}
\end{center}
\caption{The predicted $z=12$ ultraviolet luminosity functions (ULFs) for the Normal and Void {\em Renaissance Simulations}, showing flattening at faint magnitudes $M_{1600}>-15$, which roughly corresponds to the transition from LMACHs to MCHs. At the bright end $M_{1600}<-17$, the Normal ULF is in good agreement with an extrapolation of observed ULF to $z=12$ (shaded region). Reproduced by permission of the American Astronomical Society, Figure 1 from \citep{OShea15}.}
\label{fig:7}
\end{figure}

\subsection{Reionization by the first galaxies}

\bigskip
\begin{figure}[h!]
\begin{center}
\includegraphics[width=1.0\columnwidth]{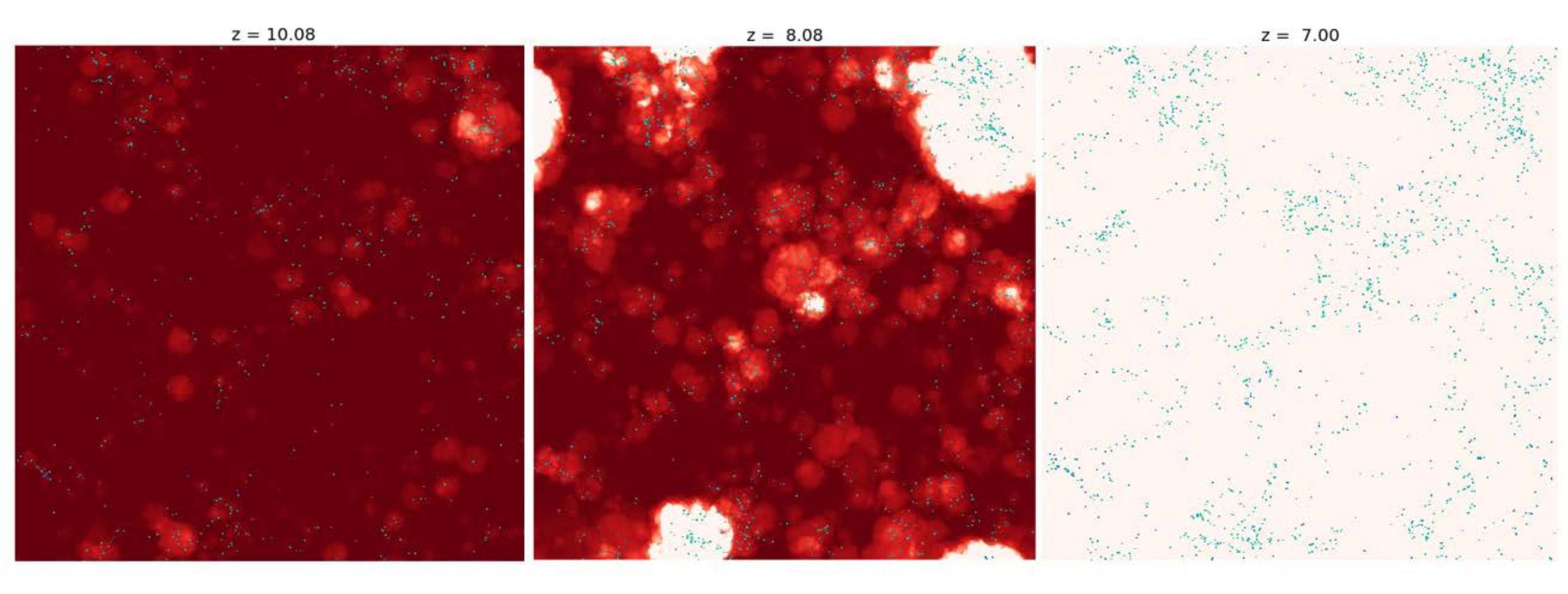}
\end{center}
\caption{An Enzo simulation of cosmic reionization with sufficient resolution to include the MCH halos discussed above. Halo finding is done inline with the simulation, and their ionizing emissivities are assigned based on results from the {\em Renaissance Simulations}. The intermittency of star formation in such halos results in stochastic early reionization. Adapted from \cite{Norman2018}.}
\label{fig:8}
\end{figure}

The interest in the shape of the faint end of the ULF stems from the fact that a substantial population of galaxies fainter than those that HST can detect are required to complete reionization by $z=7$, the latest best estimate for completion \citep{2013ApJ...768...71R}. Therefore it is interesting to see whether the population of MCHs implied by the ULF shown in Fig. \ref{fig:7} can indeed contribute to reionization, as suggested by \citep{Wise14}. We have carried out such a simulation using Enzo, and confirmed that the answer is yes. We imported the galaxy properties derived from the Renaissance Simulations into a radiation hydrodynamical simulation of hydrogen reionization. Using $1152^3$ grid cells and particles in  a small volume 10 Mpc/h on a side provides sufficient resolution to sample all halos down to a limiting mass of 
$10^7 M_{\odot}$. Halo finding was run inline with the simulation at an interval of 20 Myr. Ionizing emissivities were assigned to halos according to their mass. The intermittency of star formation in low mass halos was included using a probabilistic method consistent with the ULF shown in Fig. \ref{fig:7}. The result is shown in Fig. \ref{fig:8}. The intermittency of star formation in such halos results in stochastic early reionization. However, more massive, luminous galaxies which form stars continuously dominate below redshift 10, and complete reionization around redshift 7. Because the ionizing emissivities assigned to halos were measured at the virial radius in the Renaissance Simulations--a scale well resolved here--there was no need to assume an ionizing escape fraction, as is done is other simulations of reionization. In a real sense, the simulation shown above had no adjustable parameters, save cosmological parameters. It is reassuring that the galaxies from the Renaissance Simulations both match the observed ULF where there is overlap, and succeed in reionizing the universe by $z=7$, consistent with other observations. 

\section{The Rich Interplay between Simulations and Analytic/Semi-Analytic Models}
\label{sec:interplay}
Complementing numerical simulations are analytic and semi-analytic models of the Cosmic Dawn. There is a rich interplay between these methodologies, which we briefly review here. The literature is too large for a comprehensive review; rather, we seek to illustrate how these approaches inform one another in the context of Cosmic Dawn topics of interest. An example of an analytic model is the use of the Press-Schechter formalism \citep{PS74} to compute the halo population at high redshift, which is coupled with a parameterized model of how luminous objects form within these halos; e.g., \cite{HaimanLoeb1997,Wise05}. An example of a semi-analytic model is one that post-processes halo catalogs or merger trees from N-body dark matter simulations, adding in baryon physics through physically motivated prescriptions; e.g., \citep{Somerville2008}. Generally speaking, semi-analytic models are supplanting analytic models due to the tremendous advances in dark matter-only N-body simulations and the public availability of their data products; e.g., \citep{mill_sim,DarkSkyDR}. 

\subsection{The first stars and their feedbacks}

Once the minimum halo mass capable of hosting Pop III stars was determined analytically \citep{Tegmark97} and numerically \citep{ABN02,Bromm02_P3}, two important questions came to the fore: 1) how many Pop III stars are there globally? (Pop III star formation rate density); 2) and how long do they continue to form (duration of the Pop III epoch)? It was soon realized that Pop III stars, through their copious production of UV radiation, could have a negative feedback effect on their own formation. If Pop III stars formed in every minihalo in the universe (and why wouldn't they?), they would quickly build up a photodissociating UV background (the Lyman-Werner background) that would inhibit subsequent $H_2$ formation and cooling in minihalos. \cite{Machacek01} used Enzo simulations to show that the level of the photodissociating background sets a minimum halo mass capable of cooling. This suggested that Pop III star formation would become self-regulated globally. This was demonstrated by \cite{yoshida03}, who constructed an analytic model of Pop III star formation informed by their own numerical simulations. They calculated the Pop III SFRD for two assumptions about the Pop III stellar mass (which is still uncertain), and concluded that Pop III star formation becomes self-regulated by $z\approx 30$. 

\cite{Wise05} developed an analytic model based on the Press-Schechter formalism including self-regulated Pop III star formation in order to estimate how many Pop III supernovae might be observable. A byproduct of that is a model of the evolution of the LWB, that can be used as input to numerical simulations. The Renaissance Simulations described in Sec. \ref{subsec:rs} used an updated version of the \cite{Wise05} LWB to account for physics occurring on smaller and larger scales than could be directly simulated. Such models show that Pop III stars make only a minor contribution of the UV photons required to reionize the universe \citep{HaimanBryan2006}.    

\subsection{Chemical enrichment by the first stars and the transition to metal enriched star formation}

Regarding the second question posed above, chemical pollution of the pristine gas by stellar evolution of any kind would terminate further Pop III star formation, and usher in the formation of metal-enriched stars (Sec. \ref{subsec:pop2}). Although this process can be simulated locally, global models are needed to accurately estimate the duration of the Pop III epoch. The first attempt of this sort was by \cite{Schneider2006}, who constructed a semi-analytic model of chemical enrichment by Pop III stars. They found that unless every pristine minihalo was internally enriched by a Pop III supernova, Pop III star formation would continue to redshifts as low as z=2.5 since chemical feedback is highly localized. Hydrodynamic cosmological simulations by \cite{Tornatore2007} bore this out, and indicated that Pop III stars form preferentially at the periphery of metal-enriched zones. Analytic and semi-analytic models by \cite{Trenti2009a,Trenti2009b} confirmed and refined these estimates. Our own Renaissance Simulations, which track the chemical enrichment of the IGM with high fidelity confirm this basic view. \cite{Xu16a} carried the Void simulation to $z=7.6$ and found 14 small Pop III galaxies in the 200 cMpc$^3$ survey volume. 

To conclude this subsection we draw attention to the recent work of \cite{Visbal2018} who use high resolution catalogs from N-body simulations to construct a semi-analytic model of the transition from Pop III to metal-enriched star formation. It includes self-regulated Pop III star formation by both the LWB and local sources in the volume, and uses analytic models of metal-enriched bubbles from supernovae to determine which neighboring halos are engulfed and enriched above the threshold for Pop II star formation. In many respects, it is like the Renaissance Simulations, minus the time-consuming radiation hydrodynamics. Its principal outputs are SFRDs versus redshift for both Pop III and metal-enriched stars, with the ability to test dependencies on key input parameters. Models like this import insights from numerical simulations, and save numerical simulations from having to do expensive parameter surveys. 

\subsection{The formation of high redshift galaxies and quasars}

Semi-analytic models of galaxy formation are a mainstay of numerical cosmology to understand galaxy evolution in the post-reionization universe; e.g., \cite{mill_sim}. Increasingly, this approach is being applied to high redshift galaxy evolution in the Cosmic Dawn era. The most ambitious project of this type is DRAGONS (Dark-ages Reionization and Galaxy-formation Observables from
Numerical Simulations; \cite{Poole2016}). Over a dozen papers have been published or submitted by the DRAGONS consortium covering many of the topics we have already touched on, including the dynamical state of high redshift galaxies \citep{Poole2016}, their UV luminosity function \citep{Liu2016}, and their spatial clustering \cite{Park2017}. Such semi-analytic models require redshift outputs with fine time resolution so that accurate merger trees can be constructed. Additionally, the semi-analytic recipes need to be modified to take into account processes which are ignorable at lower redshifts. Direct comparison with hydrodynamic simulations are indispensible to this end \citep{Mitchell2018}. 

\subsection{Reionization}

Semi-analytic approaches to modeling cosmic reionization are attractive because it is a global phenomenon amenable to simplification. \cite{Choudhury2005} extended the general model of \cite{MHR00} to include radiation sources from Pop III and Pop II stars, as well as QSOs. The ionization state and thermal state of the gas is modeled in a spatially averaged way, which allows the entire dynamical model to be embedded within an optimization loop subject to observational constraints. Inputs from numerical simulations include the minimum halo mass to cool, star formation efficiencies, and the size of \hii~ regions inflated by Pop III stars in minihalos. 

\cite{FZH04} introduced a model developed along similar lines to the above that took large scale density inhomogeneities into account. They calculated the evolution of ionized \hii~ bubbles in a statistically averaged way. They showed that the characteristic bubble radius at overlap grows to $\geq 10$ cMpc \citep{Furlanetto2006}. Subsequent enhancements of this approach by Furlanetto and his colleagues have made it a versatile tool for exploring many aspects of the IGM during Cosmic Dawn. The characteristic bubble size at overlap stimulated efforts to simulate this process directly \citep{Iliev07,Norman15}. The DRAGONS semi-analytic model discussed above has also been applied to model large-scale reionization topology and to identify observational probes connected to the galaxy population \citep{Geil2016}. A key unsettled question is the relative contribution of galaxies of difference size to the reionization photon budget. 

\subsection{21cm cosmology}

Observations of the high redshift 21cm signal will probe the earliest phases of Cosmic Dawn. These observations will be particularly sensitive to the thermal evolution of the IGM in the redshift interval that is coeval with early Pop III star formation. Sophisticated global semi-analytic models have been developed by Visbal, Barkana, Fialkov, and collaborators, as recently reviewed by \cite{Barkana2016}. Key inputs from numerical simulations include the effect of the streaming velocity effect on suppressing Pop III star formation in the smallest minihalos (see Sec. \ref{subsec:streaming}), the nature of the transition from Pop III to Pop II star formation, ionizing escape fractions in high redshift galaxies, and the potential formation of X-ray sources in such galaxies. As an example of the interplay between semi-analytic models and numerical simulations, \cite{2014Natur.506..197F} emphasized the importance of high-mass X-ray binaries (HMXB) produced by the first stars and galaxies to the 21cm signal. This stimulated us to consider the possible formation of Pop III HMXB, since many Pop III stars are expected to occur in binary systems. \cite{Xu14} calculated the expected X-ray emission from galaxies in the Rare Peak Renaissance Simulation, assuming that half of all Pop III remnants were HMXBs. They showed that the preheating and preionization of the IGM is significant in such clustered regions. Moreover, \cite{Xu16c} showed that because Pop III stars continue to form at a low rate to at least the epoch of reionization, their X-ray emitting remnants alone would build up an X-ray background of considerable intensity, one that could potentially alter the temperature of the IGM in the neutral regions between \hii~ bubbles.      

\section{Physical Complications and Future Simulations}
\label{sec:complications}
\subsection{Fragmentation in primordial star formation}
\label{subsec:frag}
As discussed above, a number of works have shown the accretion disk around a Pop III protostar to be prone to fragmentation \citep{2011Sci...331.1040C, 2012MNRAS.424..399G}.  While it seems that a majority these fragments are re-accreted onto the primary object, a fraction may be ejected from the system to become low mass Pop III stars.  These calculations have been limited by the extreme computational expense of following this and further stages in the evolution of the protostar.  At the densities where fragmentation has been observed ($n > 10^{19}$ cm$^{-3}$), chemical and hydrodynamic timescales become extremely short and the core is optically thick to its own cooling radiation.  Later works have carried the torch forward by making sacrifices elsewhere.  \citet{2014ApJ...781...60H} introduced a method where Pop III-forming halos are excised from cosmological simulations and mapped into axisymmetric coordinates to be further evolved with a 2D radiation-hydrodynamics code.  \citet{2016ApJ...824..119H} improved on this with 3D rad-hydro simulations in spherical coordinates coupled to a stellar evolution code to model radiative output from the central zone.  This is a significant advance, but as of the publication of that work, the simulations had not reached the resolution required for convergence.  Additionally, the spherical geometry limits this approach to only following a single protostar and so cannot capture the evolution of any surviving clumps.

\subsection{Chemical enrichment}
\label{subsec:chemical}
The \textit{PopIIPrime} simulation showed that the external enrichment mechanism is a viable scenario for the formation of very low metallicity stars.  In particular, this mechanism is important for explaining the origins of hyper metal-poor stars like SMSS J031300.36-670839.3 \citep{2014Natur.506..463K}, which has an iron abundance of less that 10$^{-7}$ of the solar value and appears consistent with enrichment from a single supernova.  However, the frequency of such events is still not known.  As well, what are the rates and eras of relevance for other mechanisms, like self-enrichment and the accretion of pre-enriched IGM during halo assembly?  The Renaissance Simulations can provide a large enough sample from which to make statistical measurements, but their resolution is not high enough to capture the smallest Pop III minihalos of about $10^{5} M_{\odot}$.  This leaves two challenges: 1) for larger simulations at the resolution of \textit{Pop2Prime} that run past the first metal-enriched star, and 2) for a larger pool of researchers to mine the wealth of data available in works like the Renaissance Simulations.   The response to the first is a relatively straightforward technical challenge for software and hardware. The second challenge is motivated by the large volume of complex data produced by cutting-edge simulations. Data releases by the consortia that run the simulations is becoming more commonplace (see Sec. \ref{subsec:datasharing}).

\subsection{Radiation backgrounds}
Another physical complication is the role of X-rays from accreting compact objects (neutron stars, black holes) produced by the first stars and galaxies on the early intergalactic medium. Because X-rays have long mean free paths, an early population of X-ray sources will build up an X-ray background that will preionize and preheat the IGM, potentially modifying its ability to cool and form stars \citep{Machacek2003,RicottiOstriker2004,KuhlenMadau2005}. The major uncertainty is the source population. In the modern universe there are three astrophysical sources of X-rays that are well studied: active galactic nuclei, X-ray binary systems, and supernova remnants. Their relative importance as sources of X-rays in the early universe is poorly constrained by observations at the present time \citep{Fialkov2017}, leaving this a topic for speculation and simulation. One interesting possibility explored by Hao Xu and collaborators is that if some significant fraction of Pop III stars form in binary systems, as appears to be indicated by simulations (cf. Sec. \ref{subsec:pop3}), they will evolve to form high mass X-ray binary systems (HMXB). \citet{Xu14,Xu16c} showed that because Pop III star formation begins at high redshift and is ongoing to at least $z=7.6$, a significant hard X-ray background ($E_{\gamma} > 1$ keV) is produced and redshifts to lower energies where it is absorbed and heats the IGM. \citet{Xu16c} estimated the additional X-ray heating could be several hundred K by $z=6$. Future high redshift 21cm observations will probe the thermal history of the IGM during the Cosmic Dawn, thereby helping to constrain the X-ray source population. Interpreting the observations will require detailed numerical simulations complementing semi-analytic approaches \citep{Furlanetto2009,Fialkov2014}.    

\subsection{Magnetic fields}
The theoretical literature on magnetic fields in the young universe is vast, and beyond the scope of this review. Here we focus on one specific topic that has received some attention, and for which some consensus has emerged. Namely, the role of magnetic fields in the formation of the first stars. The earliest simulations of Pop III star formation ignored the role, indeed the presence, of magnetic fields entirely. This stands in contrast to modern day star formation theory, where magnetic fields have been argued to have an important if not dominant role \citep{ShuARAA,McKeeOstriker2007}. The rationale for omitting magnetic fields was not simply one of convenience. It was argued that because Pop III star formation started from pristine initial conditions, unperturbed by previous generations of star formation and other astrophysical processes, one could ignore them because they were not present. However, this overlooked the possibility that minute seed fields pervading the universe from some early magneto-genesis process could be amplified by turbulence in the protostellar cloud, potentially modifying the dynamics of collapse and fragmentation. Moreover \cite{Kulsrud1997} argued that even in the absence of primordial seed fields, structure formation itself will generate seed fields via the Biermann battery mechanism, and that these will be quickly amplified to equipartition with the turbulent kinetic energy in the protogalactic cloud. 

To investigate this possibility, \cite{Xu2008} carried out the first self-consistent AMR MHD simulations of Pop III star formation including the Biermann battery source term. The simulations were performed with the Enzo code, and were similar in design and dynamic range to the earlier simulations of \cite{ABN02} and \cite{oshea07a}. Xu et al. found that from an initial state with no magnetic fields, a combination of the Biermann battery and compressional amplification
can result in fields with strengths of $10^{-9}$ G at $n =10^{10}$ cm$^{-3}$ at the center of a cosmological halo where a Pop III star will form. At these strengths, the B-field are dynamically unimportant. As this was the highest density obtained in the simulation, the dynamical importance of B-fields during later stages of collapse could not be addressed. In parallel investigations \cite{Banerjee2012} and \cite{Turk2012} showed that the Xu et al. simulations had inadequate spatial resolution to adequately capture the turbulence in the halo's core, and that when higher resolution was used, substantially higher peak magnetic field strengths resulted. They showed that a minimum resolution threshold is required to show any turbulent amplification at all, and that above this threshold the amplification growth rate is resolution-dependent. This motivated \cite{Schober2012} to approach the issue theoretically. Using the Kazantsev theory, which
describes the so-called small-scale dynamo, they calculated the growth rate of the small-scale magnetic field for conditions appropriate to primordial halos. They confirmed what previous simulations had indicated: that assuming a seed field provided by the Biermann battery mechanism, small scale fields are amplified rapidly and can become dynamically important locally. They state that such fields are likely to become relevant after the formation of a protostellar disk and, thus, could influence the formation of the first stars and galaxies in the universe. Given what was said above about the uncertain fragmentation of the disk, this adds another layer of complexity to the subject, and calls out for detailed numerical investigation. 

\subsection{Dust}
\label{subsec:dust}
Dust is a complicating factor throughout astronomy with a wealth of literature and effort devoted to it.  In the context of simulations of the high redshift universe, dust plays an important role in chemistry and cooling of low metallicity gas and is a source of opacity for stellar radiation.  If the dust-to-gas ratio scales with the gas-phase metallicity, then the presence of dust can significantly enhance the formation of $H_{2}$ in star forming gas for metallicities as low as $10^{-4} Z_{\odot}$ and induce fragmentation in collapsing gas down to $10^{-5.5} Z_{\odot}$  \citep{2000ApJ...534..809O}.  However, there is room for significant improvement in the modeling of dust at high redshift by taking into account dust production and size distributions from Pop III supernova models \citep[e.g.,][]{2012MNRAS.419.1566S} as well as dynamic evolution of the grain population \citep{2015MNRAS.446.2659C}, just as two examples.  One of the challenges presented by improved modeling of dust grains is, of course,  the technical one, in that the increased complexity comes at a greater computational cost.  Though, perhaps a greater challenge is how the advances presented by works like those above can be incorporated into openly available simulation machinery and made available to research community.  Some potential solutions to this are discussed below.

\subsection{Dark matter physics}
\label{subsec:darkmatter}
The matter power spectrum on small scales, and how dark matter interacts in bound structures, has a profound influence on early structure formation and therefore when and how Cosmic Dawn begins. An excellent discussion of this is found in \cite{Greif2015}. Here we summarize the main points. The most explored alternative to $\Lambda CDM$ is the gavitino warm dark matter model (WDM), which truncates the matter power spectrum below a spatial scale determined by the particle mass. A mass below $\sim$ 1 keV is ruled out observationally as it would suppress the Lyman-$\alpha$ forest. \cite{Yoshida2003b} showed that a mass of 10 keV would entirely suppress the formation of minihalos of mass $\sim 10^6 \ms$ at high redshift, delaying the formation of Pop III stars until more massive dark matter assemblies form. \cite{OSheaNorman2006} examined this further using Enzo for a range of WDM particle masses 10 keV $\leq m_{WDM} \leq$ 50 keV, and found that Pop III stars form substantially later than the $m_{WDM} = \infty$ case via filament fragmentation. Importantly, however, they found that once primordial gas exceeds $10^5$ cm$^{-3}$ it cools and collapses identically to the CDM simulations. More recent work cited by \cite{Greif2015} have confirmed these results.

\cite{Maio2015,Dayal2017} have shown that the early suppression of minihalos in light WDM particle scenarios has a cascading effect on subsequent structure formation, reducing star formation rate densities for Pop III and II, shifting the galaxy luminosity function, and potentially delaying reionization. They argue that the entire Cosmic Dawn era is a sensitive probe of dark matter physics. A running spectral index of the primordial power spectrum that suppresses power on small scales has a similar effect \citep{Somerville2003}. \cite{Villanueva2018} have recently simulated reionization in a WDM cosmology, and caution that any conclusions on the nature of dark matter derived from
reionization observables remain model-dependent.

On the flip side, sterile neutrino dark matter may produce X-rays that enhance the formation of $H_2$ in minihalos, facilitating Pop III star formation \citep{Biermann2006}. Additionally, quintessence models with an evolving dark energy equation of state that satisfy observational constraints may have more dark matter minihalos at high redshift compared to $\Lambda CDM$ \citep{Maio2006}.

A current laboratory for testing alternative dark matter models is Local Group dwarf galaxies \citep{2017ARA&A..55..343B}. Given that it is generally believed that satellites of massive galaxies like the Milky Way are relics from the Cosmic Dawn, this may be the best place to do this. A variety of alternative dark matter models are being explored: self-interacting dark matter (SIDM; e.g. \cite{Elbert2015}), fuzzy dark matter (FDM; e.g. \cite{Hui2017}), and others. Insights obtained in the Local Group are being transferred to    the high redshift universe; e.g. \citep{Lovell2018}. With improved observations promised at both ultra-low and high redshifts, this is a stimulating time to be working in this subject.

\subsection{Supersonic streaming effect}
\label{subsec:streaming}
The final physical complication we will discuss concerns the so-called supersonic streaming effect on the earliest generations of Pop III star formation. This relatively recent development in the theory of early structure formation was stimulated by a paper by \cite{Hirata2010}, who showed that baryon acoustic oscillations (BAO) would result in a supersonic velocity offset between baryons and dark matter sufficient to alter how gas accretes into minihalos and possibly suppress the formation of the first bound objects that produce Pop III stars. The effect involves a quadratic term in the cosmological perturbation theory equations that was not included in studies based on linear perturbation theory. They showed that the effect leads to qualitative changes in the spatial distribution of the dark matter minihalos themselves, such as introducing scale-dependent bias and stochasticity. Naturally, this paper has generated a lot of interest among the high redshift structure formation community, and a cascade of follow-on investigations have resulted. As this is still a relatively new topic, many issues remain to be explored, and no broad consensus has emerged about the significance of the effect on later stages of structure formation. However, concerning the earliest stages of structure formation, some interesting results have been achieved, which can broadly be classified as small-scale effects and large-scale effects. 

\subsubsection{Small-scale effects}
The small-scale effects concern the accretion of baryons into dark matter mini-halos in the presence of a supersonic velocity offset of the baryons relative to the dark matter. This has been simulated by several groups using high resolution cosmological simulations with somewhat divergent results. \cite{Stacy11_Stream} performed SPH simulations initialized at
redshift 100 with a range of relative streaming velocities and minihalo formation redshifts, and found that the typical
streaming velocities have little effect on the gas evolution. They found that once the collapse begins, the subsequent evolution
of the gas is nearly indistinguishable from the case of no streaming, and star formation will still proceed in
the same way, with no change in the characteristic Pop III stellar masses. Conversely, \cite{Greif2011} performed a
series of higher-resolution moving-mesh calculations and show that these supersonic motions significantly influence
the virialization of the gas in minihalos and delay the formation of the first stars by $\Delta z \sim 4$. In addition, the streaming velocities increases the minimum halo mass capable of cooling by a factor of 3, which is partially responsible for the delay. They point out that becasue of the steepness of the halo mass function, that may reduce the number of minihalos capable of forming Pop III stars by an order of magnitude. Subsequent works by  \cite{Richardson2013} and \cite{Naoz2012,Naoz2013} find similar results to those of Greif et al., and conclude the effect is strongest for minihalos in the mass range $M/M_{\odot} < 10^6$ but becomes negligible above a halo mass of $10^{7} M_{\odot}$. We note in passing that a strong Lyman-Werner background suppresses $H_2$ formation and cooling over the same halo mass range, and so these effects must be considered together before we have a complete understanding of the earliest stages of Pop III star formation. This becomes a severe numerical challenge because of large scale effects, discussed next. 

\subsubsection{Large-scale effects}
The large-scale effects derive from the fact that the BAO modes which are responsible for the streaming velocities are themselves large-scale. The analysis of \cite{Hirata2010} showed that the streaming effect is negligible below several Mpc and above several 100 Mpc, but shows power everywhere in between. This means the streaming velocities are coherent on scales of a few Mpc--much larger than a minihalo's Lagrangian volume--but different in magnitude and direction on larger scales. The consequence is that the way in which minihalos form, accrete baryons, and cool to form Pop III stars is modulated on a range of scales between a few and a few hundred Mpc (comoving), with implications on formation epoch, spatial clustering, and radiative feedback to the IGM. In particular, the high redshift 21cm signal, which is sensitive to the gas temperature, may show structure on these scales due to the spatial modulation of how sources form on these scales. The most comprehensive large-scale models taking this effect into account, albeit in a semi-analytic way, are those of the group of Fialkov, Barkana, Visbal and collaborators, as recently reviewed by \cite{Barkana2016} and \cite{Fialkov2017}. Beyond the direct impact on 21cm and reionization, the streaming velocity effect may leave imprints on galaxy clustering \citep{2016PhRvD..94f3508S}, CMB B-mode polarization \citep{2012PhRvD..85d3523F}, early supermassive black hole formation \citep{2013MNRAS.435.3559T,2017MNRAS.471.4878S}, and may even have implications on the missing satellite problem \citep{2013ApJ...768...70B}. Recently, however, \cite{Ahn2016} improved on the analysis of \cite{Hirata2010} by including second-order terms involving density fluctuations as well (Tseliakhovich and Hirata performed their analysis assuming uniform density), and showed that density-velocity correlations may enhance the streaming velocity effect, and suggested that the entire topic needs to be revisited with this result in mind. It is clear that much remains to be done both numerically and analytically in this fascinating aspect of the early Cosmic Dawn. 

\section{Technical advances}
\label{sec:advances}
\subsection{Open source community code development}
\label{subsec:opensource}
The Enzo simulations highlighted in this paper has been made possible by the combined effort of researchers across the world.  The physical modules within the simulations, e.g., the chemistry solver, metal cooling, radiative transfer, star formation and feedback, have all resulted from separate projects performed by independent groups.  As an open-source code, Enzo has provided both a means for distribution of and access to the latest methods for simulation.  The value to the community is apparent in the number of citations (258 at the time of writing) the Enzo method paper \citep{Enzo} has received in just three years.  In fact, most of the latest work discussed here (e.g., the Renaissance Simulations, PopIIPrime, and Birth of a Galaxy) has used open-source software for its entire pipeline, from initial conditions (\citep[grafic,][]{Bertschinger01} and \citep[MUSIC][]{Hahn11}), to simulation (Enzo), and analysis \citep[yt,][]{yt_full_paper}.  As discussed earlier, the sophistication and complexity of physical modules and solvers continues to increase rapidly, as in for example, the treatment of dust grain physics.  Will researchers have to implement new published methods from scratch by following the relevant publications?  Will this work have to be repeated for every research group and simulation code?  When one considers that most scientific research is paid for by the public, the above appears especially undesirable.  In recent years, open-source chemistry packages, like Grackle \citep{2017MNRAS.466.2217S} and Krome \citep{2014MNRAS.439.2386G}, have begun to serve as these community repositories for astrochemical work and their utility is evidenced by citations to those papers as well.

\subsection{Open data sharing and analysis platforms}
\label{subsec:datasharing}
The simulations discussed here used tens of millions of CPU hours, a monetary value of at least a few million dollars.  In general, the scientific potential contained within them far exceeds the available person-power within the group that ran them.  The release of public data products, such as halo catalogs and merger trees from the Millennium Simulation \citep{mill_sim, 2006ASPC..351..212L} has greatly increased the scientific return on the initial investment.  More recently, large simulation efforts like Eagle \citep{2016A&C....15...72M, 2017arXiv170609899T} and Illustris \citep{2015A&C....13...12N} have expanded their public releases beyond halo catalogs to include full or partial snapshots.  The snapshots have a greater wealth of content, but also require significant computing facilities to store and process these large data sets.  Services like the yt Hub (https://hub.yt/) and the Renaissance Simulation Laboratory will take this one step forward by hosting data along side publicly available compute resources accessible through browser-based interfaces.  This will further decrease the necessity of expensive, local computing resources to scientific discovery.

\subsection{Toward exascale supercomputers and applications}
\label{subsec:exascale}
Impressive as the advances that have been made over the past 2 decades, it is clear that even more ambitious simulations will be needed to address the physical complications discussed in Sec. \ref{sec:complications} and future observations.  These complications, such as the large-scale effects of streaming velocities, result in an increasing {\em scale separation} that needs to be taken into account somehow. In addition, we have seen richer physical models might be required to deal with, e.g., the role of magnetic fields on Pop III protostellar disk fragmentation which will influence the Pop III stellar IMF. Finally, cosmological inferences rely on the ability to model large statistical samples of objects. The conclusion is that we need to continue moving down the track that we have already been on, namely toward larger and physically more comprehensive simulations. With the strong drive toward exascale computing systems happening worldwide, there will be adequate computer power to continue to meet growing simulation and data analysis needs. However, the architectures themselves are becoming increasingly complex, and the task of programming them is becoming a major problem. Here we briefly describe our effort to develop a version of Enzo for the coming exascale platforms.

The nominal exascale system will have something like a billion-way task or thread-level concurrency each running at a GigaHertz ($10^9 \times 10^9 = 10^{18}$). Systems will vary in their details as to how this will be accomplished, but are expected to begin showing up early in the next decade. Due to power constraints, each thread or task will need to operate out of a smaller amount of memory than we have been accustomed to (think memory per core). Also, due to power constraints, since it takes more energy to move data large distances in an HPC system, algorithms involving only local or nearest neighbor data will perform much better than algorithms that require global communication. The programmer's task therefore is to decompose the problem into of order 1 billion pieces of work that can be executed using mostly local or nearby data, and executed in such a way as to minimize the use of global synchronization or reductions. Ideally, each task would execute with its own locally-determined timestep when it has received all the data in needs to execute, i.e., asynchronously, as opposed to the bulk-synchronous programming paradigm prevalent in MPI codes. This approach is particularly advantageous for cosmological simulations where different regions of the universe evolve more or less separately from one another (gravity is the bug-bear).  

\begin{figure}[h!]
\begin{center}
\includegraphics[width=9cm]{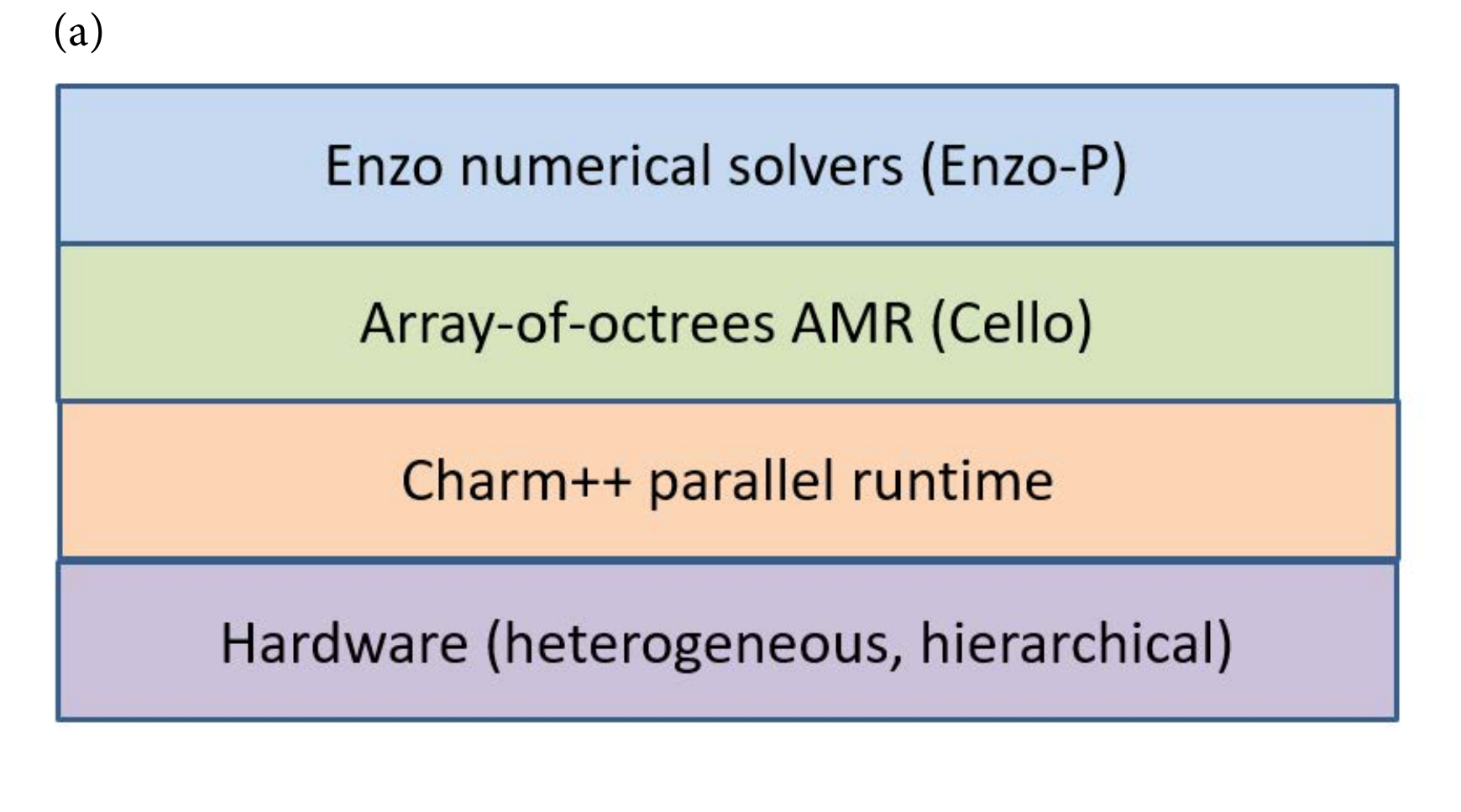}
\includegraphics[width=7cm]{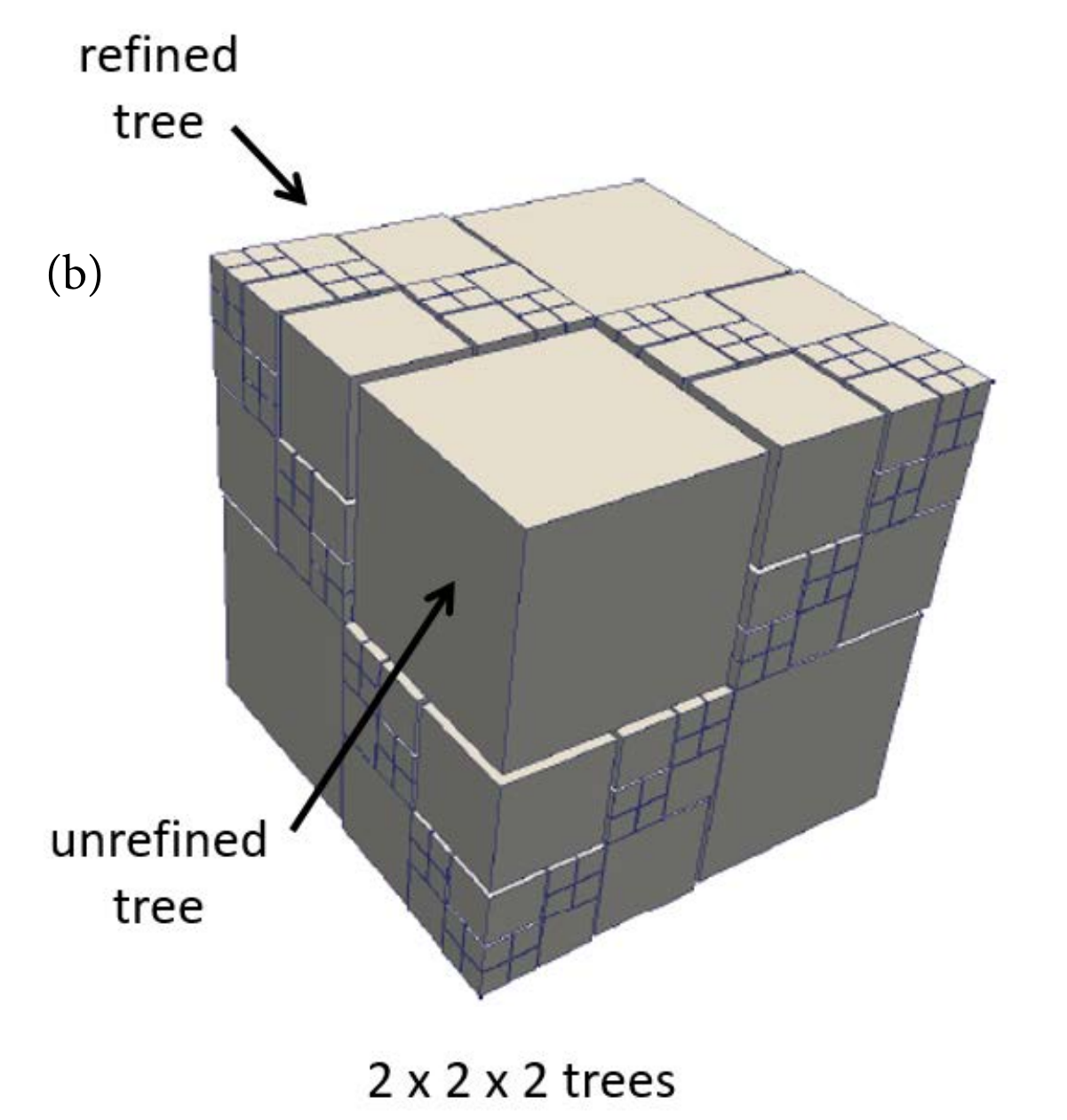}
\end{center}
\caption{(a) Layered software architecture of the Enzo-P/Cello extreme scale hydrodynamic cosmology code. (b) schematic illustration of the array-of-octree approach for adaptive mesh refinement used in Enzo-P/Cello. Each cubic block is a Cartesian grid of size $N \times N \times N$, where $N$ is user specified.}
\label{fig:12}
\end{figure}

We have designed and built such a code at UCSD called Enzo-P/Cello \citep{Bordner2018} \footnote{http://cello-project.org}. Enzo-P implements Enzo's physics solvers on top of the Cello extreme AMR software framework. Cello in turn is built on top of the Charm++ parallel object framework and runtime system developed at the University of Illinois, Urbana-Champaign. While MPI can be used to implement an asynchronous, task-parallel programming model, we chose Charm++ because it offers this capability out of the box, in addition to dynamic load balancing, fault tolerance, and parallel I/O which would otherwise fall to the application developer to implement. Charm++ is also a good fit for Enzo-P's object-oriented design. Finally, Charm++ interoperates with MPI, so libraries built with MPI can be exploited where needed. The software architecture is shown in Fig. \ref{fig:12}. We provide a thumbnail sketch of the salient features of each layer, going from bottom to top. The Charm++ parallel runtime insulates the application developer from the details of the hardware. It implements a collection of interacting objects called {\em chares} which execute user-defined tasks in a message-driven asynchronous manner. Charm++ maps the chares to processors dynamically to achieve locality and load balance. A standard aspect of Charm++ programs is the use of {\em overdecomposition} to generate many more chares (tasks) then there are processors. Overdecomposition is a good fit for exascale architectures because, as explained above, each thread or task should only operate on a small amount of data in local memory. Cello implements the array-of-octrees AMR algorithm pioneered by \cite{Burstedde2011}. As the name implies, the computational domain is covered by an array of octrees, each of which have blocks of a fixed number of cells at their leaf nodes. Fig. 12b illustrates a 3D array of size $2 \times 2 \times 2$, where some of the trees are unrefined, and some are refined by 3 additional levels. In Cello, each block is a chare, and contains local field and particle data. Chares are elements of {\em chare arrays}, which are fully-distributed data structures supported by Charm++. The index space of a chare array is user-defined and very flexible. In Cello, the chare array combines both array level indexing and tree level indexing so that a given chare can determine its position and those of its nearest neighbors using simple bit operations. As currently implemented, Cello can support a $1024^3$ array of 20-level octrees--i.e., many billions of tasks. Finally, Enzo-P is a collection of method objects which operate on the data contained in a chare (particles and fields). At the present time, Enzo-P implements most of the physics solvers in Enzo except for adaptive ray-tracing radiative transfer.

The object-oriented design of Enzo-P/Cello results in a complete separation of concerns across the 4 layers of Fig. 12. A key benefit of this is extensibility. To add a new physics capability to Enzo-P, it is as simple as adding the serial code that updates the field or particle data on a single block. AMR operations and parallel execution are handled automatically by the lower layers. Where things become somewhat more involved is implementing non-local solvers for gravity and radiation transfer which rely on linear system solvers that operate on the global adaptive mesh. In such cases Enzo-P and Cello need to be co-developed so that Cello provides the data structures and primitive operations Enzo-P methods access. 

Weak scaling tests of Enzo-P/Cello performed on the Blue Waters sustained petascale system at the National Center for Supercomputing Applications (NCSA), University of Illinois have exhibited near-ideal scaling to $64^3 = 262$K cores on a hydrodynamic AMR test problem \citep{Bordner2018}. The largest simulation performed involved an array of $64^3$ octrees, each with some 6.5M grid cells, for a total of 1.7T cells. We have also carried out strong and weak scaling experiments on a hydrodynamic cosmology test problem without AMR, which exhibits excellent scaling to 128K cores.

At the present time, Enzo-P is still under development. As a pure AMR hydrodynamics or MHD code, it is ready to use now. It is vastly superior to Enzo in terms of scalability. Cosmological hydrodynamics on a uniform mesh is also ready to use now, which exhibits superior weak and strong scaling compared to Enzo. The critical issue is the AMR gravity solver, which currently is a multigrid-preconditioned conjugate gradient algorithm. It is implemented and functioning, but needs additional tuning before it is ready for general use. We are considering implementing other approaches for gravity, such as the Fast Multipole Method, which have excellent scaling properties \citep{JabbarYK14}. Our goal is to have a feature compatible version of Enzo-P, with all of Enzo's capabilities, in 2 years time.

\section*{Author Contributions}


M. Norman wrote the majority of the review based on published results with collaborators. B. Smith wrote sections \ref{subsec:pop2}, \ref{subsec:frag}, \ref{subsec:chemical}, \ref{subsec:dust} and the first two subsections of section \ref{sec:advances} based on his published work. J. Bordner designed and developed Enzo-P/Cello described in Sec. \ref{subsec:exascale} and obtained the scaling results discussed there.

\section*{Funding}
This research has received many sources of institutional support and extramural funding over the past two decades, too many to recall. Here we acknowledge grants within the past 5 years to MLN which have been most directly connected to the first galaxies research: National Science Foundation grants AST-1109243 and AST-1615848. Enzo-P / Cello is currently funded by NSF grant SI2-SSE-1440709, with previous funding through NSF grants PHY-1104819 and AST-0808184. The PopIIPrime and Renaissance simulations were performed using ENZO
on the Blue Waters system operated by the National Center for Supercomputing
Applications (NCSA) with PRAC allocation support
by the NSF (award number ACI-0832662). Data analysis
was performed on the Comet and Gordon supercomputers operated
for XSEDE by the San Diego Supercomputer Center, and on
the Blue Waters supercomputer. 

\section*{Acknowledgments}
We gratefully acknowledge our many wonderful collaborators over the years who have contributed so much to the research presented here. In alphabetical order they are: Tom Abel, Kyungjin Ahn, Peter Anninos, Kirk Barrow, Greg Bryan, Pengfei Chen, Dave Collins, Robert Harkness, Cameron Hummels, Sadegh Khochfar, Hui Li, Shengtai Li, Dan Reynolds, Jeff Oishi, Brian O'Shea, Stephen Skory, Geoffrey So, Matt Turk, Peng Wang, Dan Whalen, John Wise, Hao Xu, and Yu Zhang. The majority of the analysis and plots were
done with YT \citep{yt_full_paper}. ENZO and YT are developed
by a large number of independent researchers from numerous
institutions around the world. Their commitment to open
science has helped make this work possible. 

\bibliography{ms}

\end{document}